\documentclass[12pt]{article}
\usepackage{amssymb,amsmath,epsfig,graphicx}
\usepackage{amsfonts}
\usepackage{amsthm}
\usepackage{caption}
\begin{document}
\title{\bf Emerging Anisotropic Compact Stars in $f(\mathcal{G},T)$ Gravity }
\author{M. Farasat Shamir \thanks{farasat.shamir@nu.edu.pk}
and Mushtaq Ahmad \thanks{mushtaq.sial@nu.edu.pk}\\\\ National University of Computer and
Emerging Sciences,\\ Lahore Campus, Pakistan.}

\date{}

\maketitle
\begin{abstract}
The possible emergence of compact stars has been investigated in the recently introduced modified Gauss-Bonnet $f(\mathcal{G},T)$ gravity, where $\mathcal{G}$ is the Gauss-Bonnet term and ${T}$ is the trace of the energy-momentum tensor \cite{sharif.ayesha}. Specifically, for this modified $f(\mathcal{G}, T)$ theory, the analytic solutions of Krori and Barua have been applied to anisotropic matter distribution. To determine the unknown constants appearing in Krori and Barua metric, the well-known three models of the compact stars namely 4U1820-30, Her X-I, and SAX J 1808.4-3658 have been used. The analysis of the physical behavior of the compact stars has been presented and the physical features like energy density and pressure, energy conditions, static equilibrium, stability,  measure of anisotropy, and regularity of the compact stars, have been discussed.
\end{abstract}


\section{Introduction}
In astrophysics, compact stars are generally being referred to as
the white dwarfs, neutron stars including the hybrid and quark
stars, and the black holes. White dwarfs and neutron stars
originated due to the degeneracy pressure produced by the
fundamental particles responsible for their formation. These stars
are massive but volumetrically smaller objects, therefore with high
densities. Usually, the exact nature of these compact stars is not
known to us but they are believed to be the massive objects with a
small radius. Excluding black holes, all the other types of compact
stars are sometimes also known as the degenerate stars. In general
relativity (GR), an analysis of configured equilibrium is essential
due to the reason of compact stars having huge mass and density.
This can be started  in a general relativistic  way by considering
the Oppenheimer-Volkoff equations \cite{OV} for static spherically
symmetric and hydrostatic equilibrium, given as
\begin{equation}\label{OV}
\frac{dp}{dr}=-G\frac{(mc^2+4p\pi{r^3})(\rho{c^2}+p)}{r^2c^4-2Gmc^2r}~~~~\text{and}~~~~\frac{dm}{dr}=4\rho{\pi}r^2,
\end{equation}
where $p$, $\rho$, and $m$ is the pressure, density, and mass of the star respectively, varying with radial coordinate $r$.
At $r=R$ ( the coordinate radius of the compact star), the total mass of compact star is determined as %For the compact star when $r=R$, the coordinate radius of the compact star, its total mass is determined as
\begin{equation}\label{OVMass}
M(R)=\int_{0}^{R}4\pi{r^2}\rho{d}{r}.
\end{equation}
The combination of these Oppenheimer-Volkoff equations with the equation of the sate (EoS) parameter $\omega=\frac{p}{\rho}$ when
solved numerically results into different configurations of the
compact stars with divisions of low and high densities \cite{Negele, AAkmal}.

Schwarzschild \cite{Schwild} was the pioneer to present the
spherically symmetric exact solutions of Einstein field equations.
The outcome of the first solution was the exploration of space-time
singularity which gave the idea of a black hole. The second non-trivial
solution predicted the bounded compactness parameter i.e.,
$\mu{(R)}=\frac{2M(R)}{R}<\frac{8}{9}$ in hydrostatic equilibrium
for some static and spherically symmetric configured structure
\cite{Buchdahl}. Investigation of compact stars (neutron stars, dark
stars, quarks, gravastars, and black holes) has become now an
interesting research pursuit in astrophysics despite being not the
new one. Baade and Zwicky \cite{Baade} studied the compact stellar
objects and argued that supernova may turn into a smaller dense
compact object which came true later on after
the discovery of pulsars which are highly magnetized rotating
neutrons \cite{Longair, Ghosh}. Ruderman \cite{Ruderman} was the first one to explore that
at the core of the compact stars, the nuclear density turns
anisotropic. A number of investigations have been made to find the
solutions of the field equations for spherically symmetric
anisotropic configurations in different contexts
\cite{Maurya}-\cite{Maharaj}. The pressure of the fluid sphere splits
into the tangential and radial pressures in anisotropic
configurations. Different investigations reveal that the repulsive
forces which construct the compact stars, are produced due to
anisotropy. Kalam et al. \cite{Kalam} showed that the Krori and
Barua  metric \cite{K&B} establishes the necessary conditions  for
the advocacy of an effective and stable approach in modelling the
compact objects. From an integrated Tolman Oppenheimer-Volkoff (TOV)
equation, the numerical simulations may be used to study the nature
of the compact stars, using EoS parameter. Rahaman et al.
\cite{Rahaman1, Rahaman2} used EoS Chaplygin gas to explore their
physical characteristics by extending Krori and
Barua models. Mak and Harko
\cite{MaK} used some standard models for spherically symmetric
compact objects and explored exact
solutions to find the physical parameters such as the energy
density, radial and tangential pressures concluding that inside
these stars, the parameters would remain positive and finite.
Hossein et al. \cite{Hossein} studied the effects on anisotropic
stars due to cosmological constant. Different physical properties such as mass, radius, and moment of inertia  of neutron stars has been investigated and a comprehensive comparison has been established with GR and modified theories of gravity \cite{32d}. Some interesting investigations related to the structure of slow rotating neutron stars in $R^2$ gravity are accomplished by making use of two distinct hadronic and a strange matter EOS parameter \cite{32e}. For a comprehensive study, some fascinating results can be seen in \cite{NJO7}-\cite{Capoz2}.

As an alternative to the
theory of  GR, modified theories of gravity have
played an important and pivotal role to reveal the hidden facts
about the accelerating expansion of the universe. After being
motivated by the original theory and using the complex lagrangian,
modified theories of gravity like $f(R)$, $f(R,T)$,
$f(\mathcal{G})$, and $f(R,\mathcal{G})$ have been structured,
where $R$ is the Ricci scalar, $\mathcal{G}$ is the Gauss-Bonnet invariant
term, and $T$ is the trace of energy momentum tensor. Some reviews and important 
discussions relating to different modified theories of
gravity have been published by different researchers
\cite{Fel}-\cite{SHZ05}. Das et al. \cite{Das} presented exact
conformal solutions to describe the interior of a star in
modified teleparallel  gravity. In another work, Das et al.
\cite{Das2} explored several physical features of the model
admitting conformal motion to describe the behavior of the compact
stars using modified $f(R,T)$ gravity. Sharif and Yousuf
\cite{SharifDas} investigated the stability conditions of collapsing
object by considering the non-static and spherically symmetric
space-time. The field equations of $f(R,T)$ modified theory have
been explored by implementing the perturbation approach \cite{Noureen3}. The possible formation of compacts stars in
 modified theory of gravity by using the Krori and Barua
metric for spherically symmetric anisotropic compact
stars has been discussed \cite{ZubairTwo, ZubairO}.
In a recently published paper \cite{sharif.ayesha}, Sharif and Ikram
presented a new modified $f(\mathcal{G},T)$ theory of gravity and
studied different energy conditions for Friedmann-Robertson-Walker
(FRW) universe. They found that the massive test particles follow
non-geodesic geometry lines due to the presence of an extra force.
It is being expected that the theory may describe the late-time
cosmic acceleration for some special choices of $f(\mathcal{G},T)$
gravity models. We discussed the Noether symmetry approach to find
the exact solutions of the field equations in $f(\mathcal{G},T)$
theory of gravity \cite{Sir&M}. In an other paper \cite{Sir&Me2}, we
investigated the same modified Gauss-Bonnet $f(\mathcal{G},T)$
gravity and used the Noether symmetry methodology to discuss some
cosmologically important $f(\mathcal{G},T)$ gravity models with
anisotropic background reported for locally rotationally symmetric
Bianchi type $I$ universe. We used two models to explore the exact
solutions and found that the
specific models of modified Gauss-Bonnet gravity may be used to
reconstruct $\Lambda$CDM cosmology without involving any
cosmological constant. Thus it seems interesting to further explore the universe in this theory.

This paper is aimed to investigate the possible emergence of compact
stars by constructing some viable stellar solutions in 
$f(\mathcal{G},T)$ theory of gravity by choosing some specific
models. The plan of our present study is as follows: In section 2, we
give the fundamental formalism of $f(\mathcal{G},T)$ gravity with
anisotropic matter distribution. Section 3 is dedicated for the
matching of the metric conditions. Some physical
features of the present study in context of $f(\mathcal{G},T)$ gravity model
under consideration are given in Section 4. Lastly, we present some
conclusive discussions.

\section{Anisotropic Matter Distribution in $f(\mathcal{G},T)$ Gravity}

The general action for the modified $f(\mathcal{G},T)$ is \cite{sharif.ayesha},
\begin{equation}\label{action}
\mathcal{A}= \frac{1}{2{\kappa}^{2}}\int d^{4}x
\sqrt{-g}[R+f(\mathcal{G},\mathrm{\textit{T}})]+\int
d^{4}x\sqrt{-g}\mathcal{L}_{M},
\end{equation}
where the function $f( \mathcal{G},T)$ consists of the Gauss-Bonnet term
$\mathcal{G}$ and the trace of the energy-momentum tensor $T$,
$\kappa$ denotes the coupling constant, $g$ is for the determinant
of the metric tensor, $R$ is the Ricci Scalar, and $\mathcal{L}_{M}$
represents matter part of the Lagrangian. The Gauss-Bonnet term $\mathcal{G}$
is defined as
\begin{equation}
\mathcal{G}=R^{2}-4R_{\zeta\eta}R^{\zeta\eta}+R_{\zeta\eta\mu\nu}R^{\zeta\eta\mu\nu},
\end{equation}
where $R_{\zeta\eta\mu\nu}$, and $R_{\zeta\eta}$ is the Reimann and
Ricci tensors, respectively. The variation of Eq.(\ref{action}) with
respect to $g_{\zeta\eta}$,  and by setting $\kappa=1$, gives the
following fourth order non-linear field equations
\begin{eqnarray}\nonumber
G_{\zeta\eta}&=&[2Rg_{\zeta\eta}\nabla^{2}+
2R\nabla_{\zeta}\nabla_{\eta}+4g_{\zeta\eta}R^{\mu\nu}\nabla_{\mu}\nabla_{\nu}+
4R_{\zeta\eta}\nabla^{2}-\\\nonumber
&&4R^{\mu}_{\zeta}\nabla_{\eta}\nabla_{\mu}-4R^{\mu}_{\eta}\nabla_{\zeta}\nabla_{\mu}-4R_{\zeta\mu\eta\nu}\nabla^{\mu}\nabla^{\nu}]f_{\mathcal{G}}+
\frac{1}{2}g_{\zeta\eta}f-[\mathrm{\textit{T}}_{\zeta\eta}+\Theta_{\zeta\eta}]\times\\
&&f_{\mathrm{\textit{T}}}-[2RR_{\zeta\eta}-4R^{\mu}_{\zeta}R_{\mu\eta}-4R_{\zeta\mu\eta\nu}R^{\mu\nu}+2R^{\mu\nu\delta}_{\zeta}R_{\eta\mu\nu\delta}]
f_{\mathcal{G}}+\kappa^{2}\mathrm{\textit{T}}_{\zeta\eta},\label{4_eqn}
\end{eqnarray}
where $\Box=\nabla^{2}=\nabla_{\zeta}\nabla^{\zeta}$ is the
d'Alembertian operator, ${G}_{\zeta\eta}=R_{\zeta\eta}-\frac{1}{2}g_{\zeta\eta}R$ is the
Einstein tensor, $\Theta_{\zeta\eta}= g^{\mu\nu}\frac{\delta
\mathrm{\textit{T}}_{\mu\nu}}{\delta g_{\zeta\eta}}$, $f\equiv f(\mathcal{G},T)$, $f_{\mathcal{G}}\equiv\frac{\partial f (\mathcal{G},\mathrm{\textit{T}})}{\partial
\mathcal{G}}$, and $f_{\mathrm{\textit{T}}}\equiv\frac{\partial
f(\mathcal{G},\mathrm{\textit{T}})}{\partial \mathrm{\textit{T}}}$.
Einstein equations can be reawakened by putting simply
$f(\mathcal{G},\mathrm{\textit{T}})=0$ whereas field equations for
$f(\mathcal{G})$  are reproduced  by replacing
$f(\mathcal{G},\mathrm{\textit{T}})$ with $f(\mathcal{G})$ in Eq.($\ref{4_eqn}$).
The energy-momentum tensor denoted by $\mathrm{\textit{T}}_{\zeta\eta}$ can be defined as
\begin{equation}\label{emt}
\mathrm{\textit{T}}_{\zeta\eta}=-\frac{2}{\sqrt{-g}}\frac{\delta(\sqrt{-g}\mathcal{L}_{M})}{\delta
g^{\zeta\eta}}.
\end{equation}
Moreover, the metric dependent energy-momentum tensor may have the form
\begin{equation}\label{emt1}
\mathrm{\textit{T}}_{\zeta\eta}=g_{\zeta\eta}\mathcal{L}_{M}-2\frac{\partial\mathcal{L}_{M}}{\partial
g^{\zeta\eta}}.
\end{equation}
The usual anisotropic energy momentum tensor $T_{\zeta\eta}$ is given as
\begin{equation}\label{12}
T_{\zeta\eta}=(\rho+p_{t})V_{\zeta}V_{\eta}-p_{t}g_{\zeta\eta}+(p_{r}-p_{t})\xi_{\zeta}\xi_{\eta},
\end{equation}
where $p_{t}$ and  $p_{r}$ represents the tangential and radial
pressures  respectively  while $\rho$ denotes the  energy density.
The four velocity is denoted by $V_{\zeta}$ and  the radial four
vector by $\xi_{\alpha}$, satisfying
\begin{equation}\label{13}
V^{\alpha}=e^{\frac{-a}{2}}\delta^{\alpha}_{0},~~~V^{\alpha}V_{\alpha}=1,~~~\xi^{\alpha}=e^{\frac{-b}{2}}\delta^{\alpha}_{1},~~~\xi^{\alpha}\xi_{\alpha}=-1.
\end{equation}
In this paper, we have chosen specifically the following  $f(\mathcal{G},T)$ model \cite{Sir&M}
\begin{equation}\label{VM}
f(\mathcal{G},T)=f_{1}(\mathcal{G})+f_{2}(T),
\end{equation}
where $f_{1}(\mathcal{G})$ is an analytic function comprised of Gauss-Bonnet
term. In particular, we consider
$f_{1}(\mathcal{G})=\alpha\mathcal{G}^n$, a power law model of
$f(\mathcal{G})$ gravity proposed by Cognola et al. \cite{17} with
$\alpha$ being an arbitrary real constant, and $n$ a positive real
number. Here we take $f_{2}({T})=\lambda\textit{T}$, with $\lambda$
being some positive real number. Further, for the investigations on
the compact stars we take the static, spherically symmetric
space-time as
\begin{equation}\label{11}
ds^{2}=e^{a(r)}dt^{2}-e^{b(r)}dr^2-r^{2}d\theta^{2}-r^2sin^{2}\theta d\phi^{2}.
\end{equation}
We parameterize metric (\ref{11}) by taking $a(r)=Br^{2}+C$ and $b(r)=Ar^{2}$, given by Krori and Barua \cite{K&B} and with the help of some physical assumptions, the arbitrary constants $A, B$ and $C$ will be calculated. The above set of functions are established to reach a singularity free structure for compact stars. Therefore, our main concern is to present these functions to the metric in a way to achieve the structure for the compact star under this extended model, free from the singularities in the neighbourhood of $r$.
To investigate the existence of the compact stars for the model $f(\mathcal{G},T)=\alpha\mathcal{G}^{n}+\lambda\textit{T}$, we have considered $\alpha=1$, $\lambda=2$, and $n=2$. For these parametric values, the energy density and all the energy conditions remain positive for the model under investigation. It is worth mentioning here that one may opt for some other choices of these values for further analysis.
The explicit expressions for the energy density $\rho$, radial pressure $p_r$, and the tangential pressure $p_t$ are obtained as
\begin{eqnarray}\label{Rho1}
\rho&=&\frac{e^{-2b}}{8r^4(1+\lambda)(1+2\lambda)}\Big[8e^b(-1+e^b)r^2(1+2\lambda)-4e^{2b}r^4(1+\lambda)f_1\\\nonumber
&+&f_\mathcal{G}\Big\{-16(-1+e^b)^2\lambda+r^2\Big\{r^2(1+2\lambda)a'^4-2r^2(1+2\lambda)a'^3b'\\\nonumber
&-&4a'b'\Big\{2(-3+e^b)(1+\lambda)+r^2(1+2\lambda)a''+a'^2\Big\{8(\lambda+e^b(1+\lambda))\Big\}\\\nonumber
&+&r^2(1+2\lambda)(b'^2+4a'')\Big\}+4\Big\{-2\lambda
b'^2+a''\{4(-1+e^b)(1+\lambda)\}\Big\}\\\nonumber
&+&r^2(1+2\lambda)a''\Big\}\Big\}+2r\Big\{4f_\mathcal{G}'\{-8(2+5\lambda)+r\{b'(10+27\lambda-2r\lambda
b')\}\\\nonumber
&-&r\{8+18\lambda+r(2+3\lambda)b'\}a''\}-8r(2+5\lambda)(1-2rb'+r^2a'')f_\mathcal{G}''\\\nonumber
&+&r^2a'^2\Big\{e^{b}r\lambda-2(8+18\lambda+r(2+3\lambda)b')f_\mathcal{G}'-4r(2+5\lambda)f_\mathcal{G}''\Big\}\\\nonumber
&+&2e^b\Big\{16(2+5\lambda)f_\mathcal{G}'+2rb'\{r+2r\lambda+(2+3\lambda)f_\mathcal{G}'\}+r^3\lambda
a''+4r(2+5\lambda)f_\mathcal{G}''\Big\}\\\nonumber
&+&ra'\Big\{2(-32-74\lambda+rb'(2\lambda+r(2+3\lambda)b'))\Big\}f_\mathcal{G}'-e^b\lambda\Big\{r(-4+r
b')+4f_\mathcal{G}'\Big\}\\\nonumber
&+&4r\Big\{-2(4+9\lambda)+r(2+5\lambda)b'f_\mathcal{G}''\Big\}\Big\}\Big],
\end{eqnarray}
\begin{eqnarray}\label{pr1}
p_r&=&\frac{e^{-2b}}{8r^4(1+\lambda)(1+2\lambda)}\Big[-4e^{2b}r^4(1+\lambda)f_{1}+f_\mathcal{G}\Big\{-16(-1+e^b)^2\lambda\\\nonumber
&+&r^2\Big(r^2(1+2\lambda)a'^4-2r^2(1+2\lambda)a'^{3}b'-4a'b'\{2(-3+e^b)(1+\lambda)\\\nonumber
&+&r^2(1+2\lambda)a''\}+a'^2\Big\{8(-1+e^b)(1+\lambda)+r^2(1+2\lambda)(b'^2+4a'')\Big\}\\\nonumber
&+&4\Big\{2(1+\lambda)b'^2+a''\{4(-1+e^b)(1+\lambda)+r^2(1+2\lambda)a''\}\Big\}\Big)\Big\}\\\nonumber
&+&2r\{4e^b(-1+e^b)\}r(1+2\lambda)+4f_\mathcal{G}'\Big\{-8\lambda+r\Big\{-b'(4+\lambda+2r\lambda
b')\\\nonumber &+&r\lambda(-2+r b'a'')+r^2\lambda
a'^2\{2(-2+rb')f_\mathcal{G}'+r(e^b-4f_\mathcal{G}'')\}-8r\lambda(1-2rb'\\\nonumber
&+&r^2a'')f_\mathcal{G}''+ra'\Big\{-2\{12+34\lambda+rb'(8+14\lambda+r\lambda
b')\}f_\mathcal{G}'\\\nonumber &+&e^b\Big\{-r(4+4\lambda+r\lambda
b')+4(2+3\lambda)f_\mathcal{G}'\Big\}+4r\lambda(-2+rb')f_\mathcal{G}''\Big\}\\\nonumber
&+&2e^b[2\{8\lambda+r(4+7\lambda)b'\}f_\mathcal{G}'+r\lambda(r^2a''+4f_\mathcal{G}'')]\Big\}\Big\}\Big],
\end{eqnarray}
\begin{eqnarray}\label{pt1}
p_t&=&\frac{e^{-2b}}{4r^4(1+\lambda)(1+2\lambda)}\Big[-2e^{2b}r^4(1+\lambda)f_1+2f_\mathcal{G}\Big\{4(-1+e^b)^2(1+\lambda)\\\nonumber
&+&r^2\Big\{\{-1+2e^b(1+\lambda)\}a'^2-2(-3+e^b)(1+\lambda)a'b'+b'^2+4(-1+e^b)(1\\\nonumber
&+&\lambda)a''\Big\}\Big\}-r\Big\{4f_\mathcal{G}'[8\lambda+r\{b'(-7\lambda-2r(1+\lambda)b')+r(2+6\lambda\\\nonumber
&+&r(2+3\lambda)b')a''\}]+8r\lambda(1-2rb'+r^2a'')f_\mathcal{G}''+2e^b\{-16\lambda
f_\mathcal{G}'-rb'(r+2r\lambda\\\nonumber &-&2\lambda
f_\mathcal{G}')+r^3(1+\lambda)a''-4r\lambda
f_\mathcal{G}''\}+r^2a'^2\Big\{e^br(1+\lambda)+2\Big\{2+6\lambda\\\nonumber
&+&r(2+3\lambda)b'\Big\}f_\mathcal{G}'+4r\lambda
f_\mathcal{G}''\Big\}+ra'-2[-10\lambda+rb'\{2+6\lambda\\\nonumber
&+&r(2+3\lambda)b'\}]f_\mathcal{G}'+e^b[-r\{-2+r(1+\lambda)b'\}+4\lambda
f_\mathcal{G}']+4r(2+6\lambda-r\lambda
b')f_\mathcal{G}''\Big\}\Big],
\end{eqnarray}
where prime denotes the radial derivative.

\section{Matching With Schwarzschild's Exterior Metric}

Whatever the geometry of the star is, either derived internally or
externally, the intrinsic boundary metric remains the same. Thus,
confirming that the components of the metric tensor irrespective of
the coordinate system  across the surface of the boundary will
remain continuous. No doubt, in GR, the Schwarzschild solutions have been pioneer in guiding us to choose from the diverse possibilities of the matching conditions  while investigating the stellar compact objects. Now when we come to the case of modified theories of gravity, modified TOV equations with zero
pressure and energy density, the solution outside the star can differ from Schwarzschild's solution. However,
it is expected that the solutions of the modified TOV equations with energy density and pressure (may be non-zero) may accommodate Schwarzschild's solution with some specific choice of $f(\mathcal{G},T)$ gravity model. Perhaps this is the reason that Birkhoff's theorem may not hold in modified gravity. The detailed investigation of the issue in the context of $f(\mathcal{G},T)$ gravity can be an interesting task. Many authors have considered Schwarzschild solution for this purpose giving some interesting results \cite{27a}-\cite{Ast}.
Now to solve the field equations under the restricted
boundary conditions at $r=R$, the pressure $p_{r}=0$, the interior
metric (\ref{11}) requires these matching  conditions. This can be
done by taking a smooth match at $r=R$ to Schwarzschild's exterior
metric, given by
\begin{equation}\label{14}
ds^{2}=\Big(1-\frac{2M}{r}\Big)dt^{2}-\Big(1-\frac{2M}{r}\Big)^{-1}dr^{2}-r^{2}(d\theta^{2}+sin^{2}\theta
d\phi^{2}),
\end{equation}
yielding
\begin{equation}\label{15}
g^{-}_{tt}=g^{+}_{\alpha},~~~g^{-}_{rr}=g^{+}_{rr},~~~\frac{\partial
g^{-}_{\alpha}}{\partial r}=\frac{\partial g^{+}_{\alpha}}{\partial
r},
\end{equation}
where $(+)$ corresponds to exterior solution and $(-)$ to the interior solution. Now from the comparison of exterior and interior metrics, the constants $A$, $B$, and $C$ are obtained as
\begin{eqnarray}
A&=&\frac{-1}{R^{2}}ln\Big(1-\frac{2M}{R}\Big),\\
B&=&\frac{M}{R^{3}}\Big(1-\frac{2M}{R}\Big),\\
C&=& ln\Big(1-\frac{2M}{R}\Big)-\frac{M}{R}\Big(1-\frac{2M}{R}\Big).
\end{eqnarray}
The approximated values of the mass $M$ and radius $R$
of the compact stars Her X-1, SAX
J1808.4-3658 and 4U1820-30 are used to calculate the values of constants $A$ and
$B$ \cite{Lattimer,Li}. The result $\mu=\frac{M(R)}{R}$ defines the compactness of the
star and the expression $Z_{s}=(1-2\mu)^{-1/2}-1$ determines the
surface redshift $Z_{s}$. For the compact stars under consideration,
the values of $Z_{s}$ have been given below in table as follows.
\begin{center}
\begin{tabular}{ |c|c|c|c|c|c|c| }
 \hline
 Compact Stars  &      $M$           & $R(km)$  & $\mu=\frac{M}{R}$& $A(km^{-2})$     & $B(km^{-2})$   & $Z_{s}$ \\
 \hline
 Her X-1&0.88$M_{\odot}$  & 7.7    &  0.168              & 0.006906276428   & 0.004267364618 & 0.23  \\
 \hline
 SAXJ1808.4-3658&1.435$M_{\odot}$ & 7.07   &  0.299        & 0.01823156974    & 0.01488011569  & 0.57  \\
 \hline
 4U1820-30&2.25$M_{\odot}$  & 10.0   &  0.332              & 0.01090644119    & 0.009880952381 & 0.73 \\
 \hline
\end{tabular}
\captionof{table}{The approximate values of the masses $M$, radii $R$, compactness $\mu$, and the constants $A$, and $B$ for the compact stars Her X-1, SAXJ 1808.4-3658, and 4U 1820-30}\label{Table:1}
\end{center}
Now making use of Krori and Barua metric, Eqs.
(\ref{Rho1})-(\ref{pt1}) take the form
\begin{eqnarray}\label{Rho2}
\rho&=&\frac{e^{-2Ar^{2}}}{2r^{4}(1+\lambda)(1+2\lambda)}\Big[4\Big\{-\lambda-e^{2Ar^{2}}\lambda+2e^{Ar^{2}}\{\lambda+Br^{2}(1\\\nonumber
&+&(-A+B)r^{2})(1+\lambda)+r^{2}\{-2A^{2}r^{2}\lambda+2B(-1+3Ar^{2})(1+\lambda)\\\nonumber
&+&B^{4}r^{6}(1+2\lambda)-2B^{3}r^{4}(-1+Ar^{2})(1+2\lambda)+B^{2}r^{2}(1+4\lambda\\\nonumber
&+&Ar^{2}(-2+Ar^{2})(1+2\lambda))\}\}\Big\}f_{\mathcal{G}}+r\Big\{-4(2+5\lambda)+r^{2}(10A-24B\\\nonumber
&-&4B(A+2B)r^{2}+4A(A-B)Br^{4}+\{27A-55B-2(2A^2+2AB\\\nonumber
&+&9B^{2})r^{2}+6A(A-B)Br^{4}\}\lambda)+e^{Ar^{2}}[8+20\lambda+r^{2}(-B\lambda+A(2\\\nonumber
&+&3\lambda))]f_{\mathcal{G}}'+r(e^{Ar^{2}}(-2+4Ar^{2}-4\lambda+2r^{2}(4A+3B+B(-A\\\nonumber
&+&B)r^{2})\lambda+e^{Ar^{2}}(2+4\lambda-8^{n}r^{2}\\\nonumber
&\times&\frac{1}{r^{2n}}\Big(Be^{-2Ar^{2}}(1+(-3A+B)r^{2}+e^{Ar^{2}}(-1+(A-B)r^{2}))\Big)^{n}\\\nonumber
&\times&\alpha(1+\lambda)))+4(-2-5\lambda+e^{Ar^{2}}(2+5\lambda)+2r^{2}(A(2+Br^{2})(2+5\lambda)\\\nonumber
&-&B(6+14\lambda+Br^{2}(2+5\lambda))))f_{\mathcal{G}}'')\Big\}\Big],
\end{eqnarray}
\begin{eqnarray}\label{pr2}\
p_{r}&=&\frac{e^{-2Ar^{2}}}{2r^{4}(1+\lambda)(1+2\lambda)}\Big[4\Big\{-(-1+e^{Ar^{2}})^{2}\lambda+2A^{2}r^{4}(1+\lambda)\\\nonumber
&-&2Br^{2}(1-3Ar^{2}+e^{Ar^{2}}(-1+Ar^{2}))(1+\lambda)+B^{4}r^{8}(1+2\lambda)\\\nonumber
&-&2B^{3}r^{6}(-1+Ar^{2})(1+2\lambda)+B^{2}r^{4}(-1+2e^{Ar^{2}}(1+\lambda)+Ar^{2}(-2\\\nonumber
&+&Ar^{2})(1+2\lambda))\Big\}f_\mathcal{G}+r(e^{Ar^{2}}r(e^{Ar^{2}}(2+4\lambda-8^{n}r^{2}\\\nonumber
&\times&\frac{1}{r^{2n}}\Big(Be^{-2Ar^{2}}(1+(-3A+B)r^{2}+e^{Ar^{2}}(-1+(A-B)r^{2}))\Big)^{n}\\\nonumber
&\times&\alpha(1+\lambda))-2(1+2\lambda+Br^{2}(2+(1+(A-B)r^{2})\lambda)))\\\nonumber
&+&4(-4\lambda-r^{2}(4A+6B+8ABr^{2}+(A+19B+2(2A^{2}+6AB+B^{2})r^{2}\\\nonumber
&+&2A(A-B)Br^{4})\lambda)+e^{Ar^{2}}(4\lambda+r^{2}(B(2+3\lambda)+A(4+7\lambda))))f_\mathcal{G}'\\\nonumber
&+&4r(-1+e^{Ar^{2}}+2(A-B)r^{2}(2+Br^{2}))\lambda
f_\mathcal{G}'')\Big],
\end{eqnarray}
\begin{eqnarray}\label{pt2}
p_{t}&=&\frac{e^{-2Ar^{2}}}{2r^{4}(1+\lambda)(1+2\lambda)}\Big[4(1+A^{2}r^{4}-B^{2}r^{4}+\lambda+e^{2Ar^{2}}(1+\lambda)\\\nonumber
&+&2Br^{2}(-1+3Ar^{2})(1+\lambda)+2e^{Ar^{2}}(-1+Br^{2}+B(-A+B)r^{4})\\\nonumber
&\times&(1+\lambda))f_\mathcal{G}+r(e^{Ar^{2}}r^{3}(-8^{n}e^{Ar^{2}}\times\frac{1}{r^{2n}}\Big(Be^{-2Ar^{2}}(1+(-3A+B)r^{2}\\\nonumber
&+&e^{Ar^{2}}(-1+(A-B)r^{2}))\Big)^{n}\alpha(1+\lambda)-2B(2+\lambda+Br^{2}(1+\lambda))+2A\\\nonumber
&\times&(1+2\lambda+Br^{2}(1+\lambda)))+4(4(-1+e^{Ar^{2}})\lambda+r^{2}(A(-(-7+e^{Ar^{2}})\lambda\\\nonumber
&+&4Ar^{2}(1+\lambda))-2B^{2}r^{2}(1+3\lambda+Ar^{2}(2+3\lambda))+B(-2-(11+e^{Ar^{2}}))\lambda\\\nonumber
&+&2Ar^{2}(-1+Ar^{2}(2+3\lambda)))))f_\mathcal{G}'-4r(\lambda-e^{Ar^{2}}\lambda+2r^{2}(B-2A\lambda\\\nonumber
&+&B(4+(-A+B)r^{2})\lambda))f_\mathcal{G}'')\Big],\\
%\end{eqnarray}
\text{where}\\\nonumber
%\begin{eqnarray}\nonumber
f_{1}&=&\alpha\Big[\frac{e^{-2Ar^{2}}
(-16ABr^{2}+2(1-e^{Ar^{2}})(4B-4ABr^{2}+4B^{2}
r^{2}))}{r^{2}}\Big]^{n},\\
%\end{eqnarray},
%\begin{eqnarray}
f_{\mathcal{G}}'&=&\frac{1}{r^{3}}\Big[2^{-2+3n}Be^{-2Ar^{2}}(-1+n)n(Be^{-2Ar^{2}}(1+(-3A+B)r^{2}\\\nonumber
&+&e^{Ar^{2}}(-1+(A-B)r^{2}))){r^{2}}^{-2+n}(-1+6A^{2}r^{4}-2A(r^{2}+Br^{4})\\\nonumber
&+&e^{Ar^{2}}(1+Ar^{2}+A(-A+B)r^{4}))\alpha\Big],
\end{eqnarray}
\begin{eqnarray}\label{pt32}
f_{\mathcal{G}}''&=&\frac{4^{-1+2n}e^{3Ar^{2}}(-1+n)n\alpha}{B(-1+(3A-B)r^{2}+e^{Ar^{2}}(1+(-A+B)r^{2}))^{3}}\\\nonumber
&\times&\frac{1}{r^{2n}}\Big(-Be^{-\frac{3Ar^{2}}{2}}(Ar^{2}cosh(\frac{Ar^{2}}{2}))+(1+(2A+B)r^{2})sinh(\frac{Ar^{2}}{2}))\Big)^{n}\\\nonumber
&\times&(2-4n-3(2B+A(-9+4n))r^{2}+2A(9B-6Bn+A(-21+10n))r^{4}\\\nonumber
&+&A(3B^2-16ABn+A^{2}(-15+32n))r^{6}-2A^{2}(3A^{2}-4AB+B^{2})\\\nonumber
&\times&(-3+4n)r^{8}+(-2+4n+3(2B+A(-9+4n))r^{2}+2A(16A-9B\\\nonumber
&-&9An+6Bn)r^{4}+A(-3B^{2}+A^{2}(33-52n)+2AB(-3+10n))r^{6}\\\nonumber
&+&2A^{2}(37A^{2}-26AB+5B^{2})(-1+n)r^{8})cosh(Ar^{2})+Ar^{2}(11-4n\\\nonumber
&+&2(B(3-2n)+A(-12+7n))r^{2}+(B^{2}+2AB(1-6n)+A^{2}(-27\\\nonumber
&+&44n))r^{4}-2A(7A-3B)(5A-B)(-1+n)r^6)sinh(Ar^2)).
\end{eqnarray}
Now we investigate the nature and some interesting features of the compact star, specifically for the assumed model of $f(\mathcal{G},T)$ gravity.

\section{Physical Aspects of $f(\mathcal{G},T)$ Gravity Model}

In this section, some interesting physical aspects of the compact
stars such as the energy density and pressure evolutions, energy
conditions, equilibrium conditions, stability and adiabatic index
analysis, compactness, and redshift analysis shall be discussed.
\subsection{Energy Density and Pressure Evolutions}
\begin{figure}\center
\begin{tabular}{cccc}
\\ &
\epsfig{file=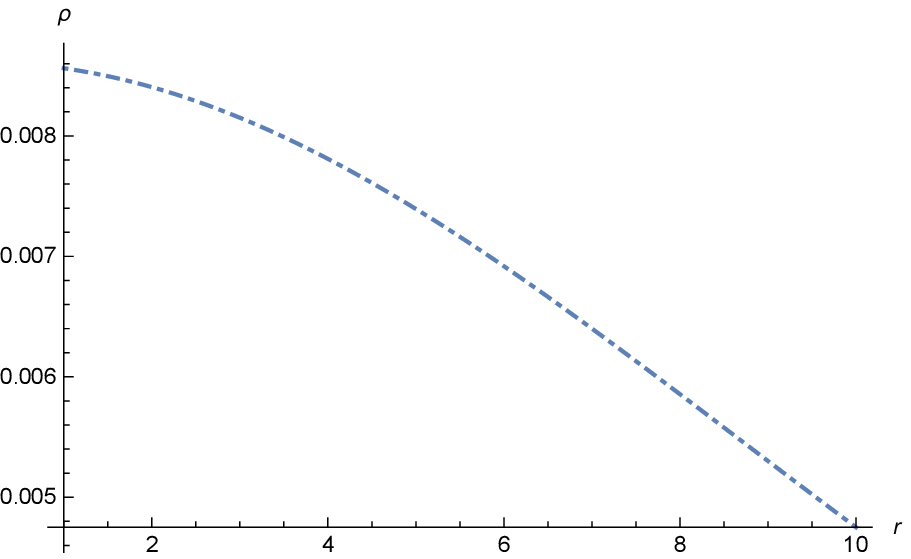,width=0.35\linewidth} &
\epsfig{file=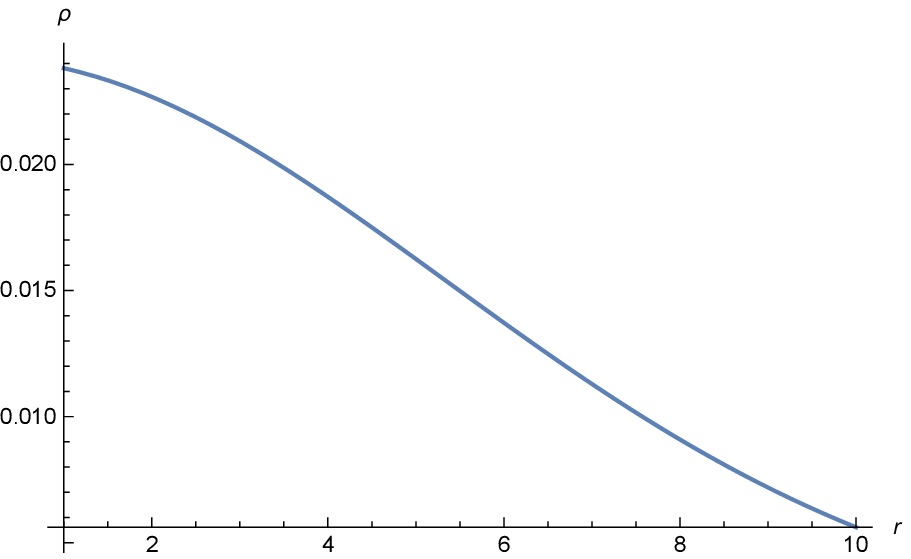,width=0.35\linewidth} &
\epsfig{file=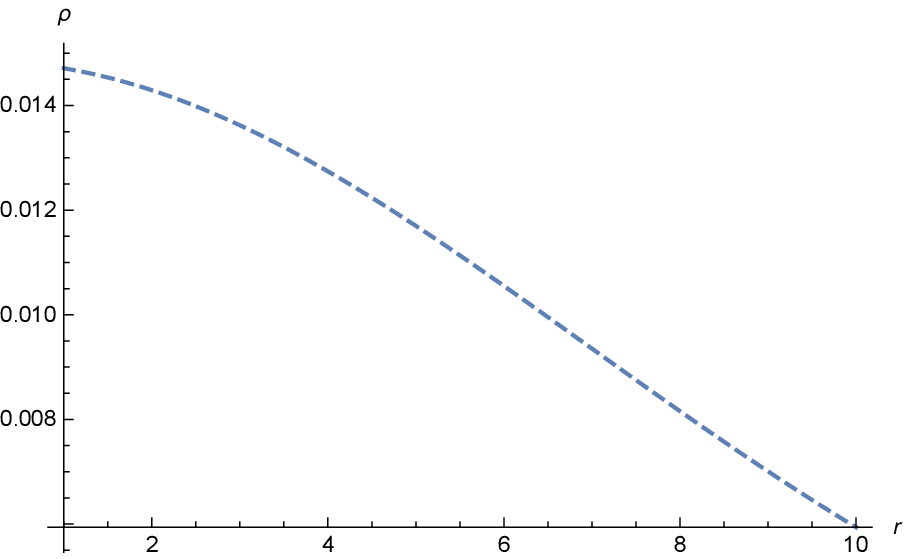,width=0.35\linewidth}\\
\end{tabular}
\caption{Plot of the density evolution of the strange star candidate
Her X-1, SAX J 1808.4-3658, and 4U 1820-30; represented by
(left to right) 1st, 2nd, and 3rd graphs,
respectively.\label{fig:Densitygraphs}}\center
\end{figure}
\begin{figure}\center
\begin{tabular}{cccc}
\\ &
\epsfig{file=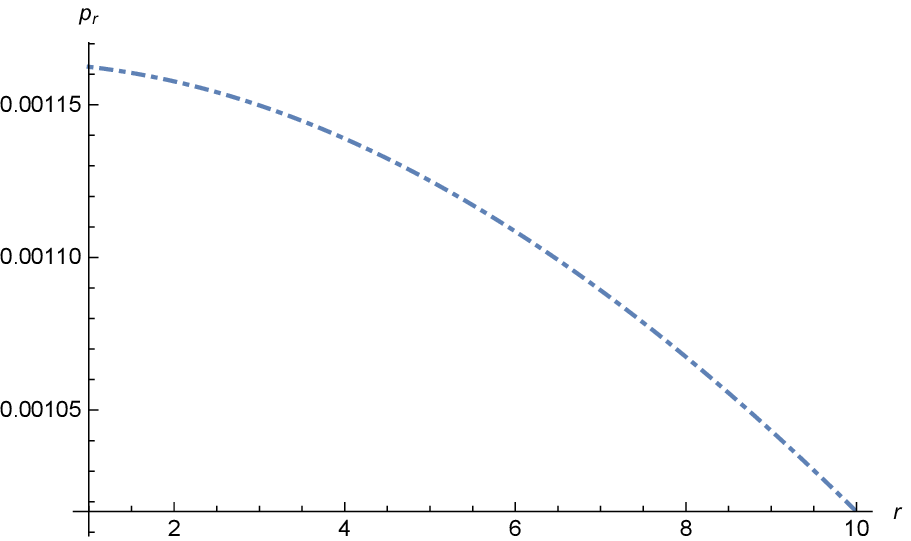,width=0.35\linewidth} &
\epsfig{file=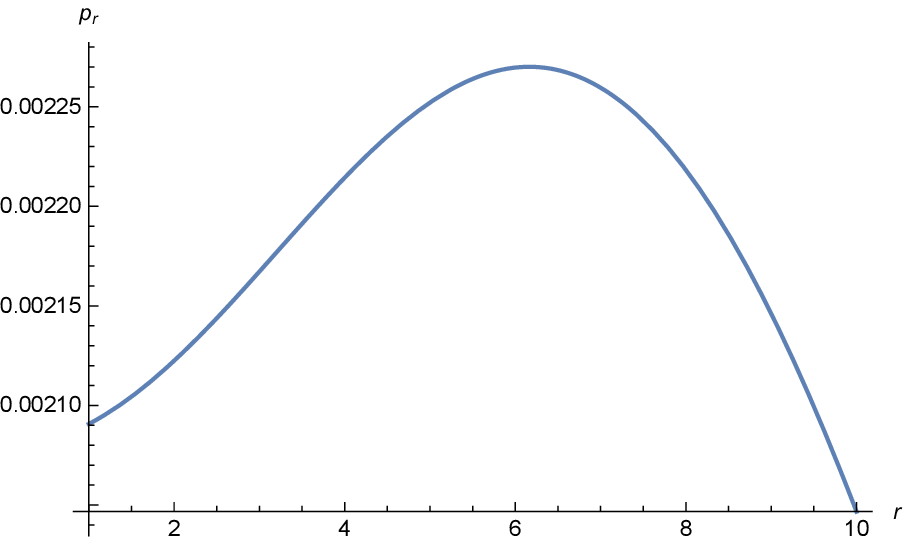,width=0.35\linewidth} &
\epsfig{file=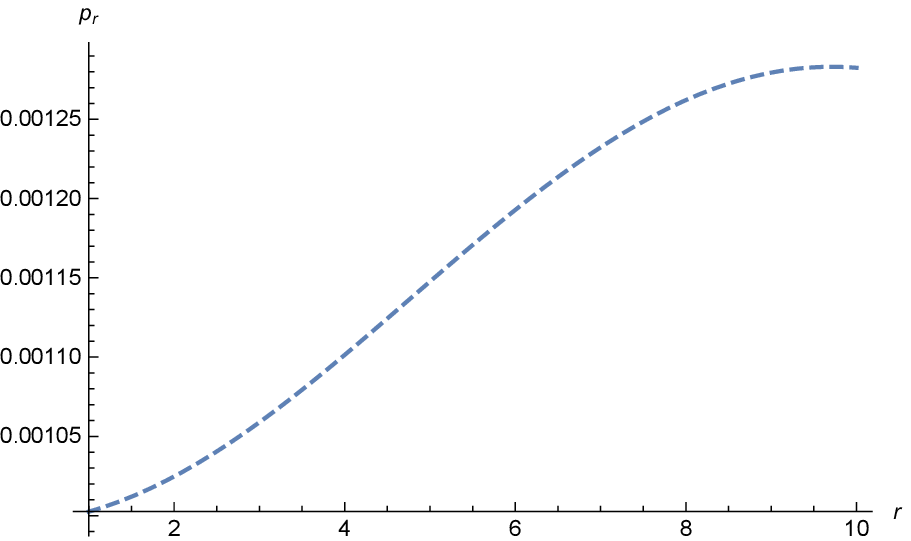,width=0.35\linewidth}\\
\end{tabular}
\caption{Plots of the radial pressure evolution of the strange star
candidates Her X-1, SAX J 1808.4-3658, and 4U 1820-30;
represented by (left to right) 1st, 2nd, and 3rd graphs,
respectively.\label{fig:prgraphs}}\center
\end{figure}
\begin{figure}\center
\begin{tabular}{cccc}
\\ &
\epsfig{file=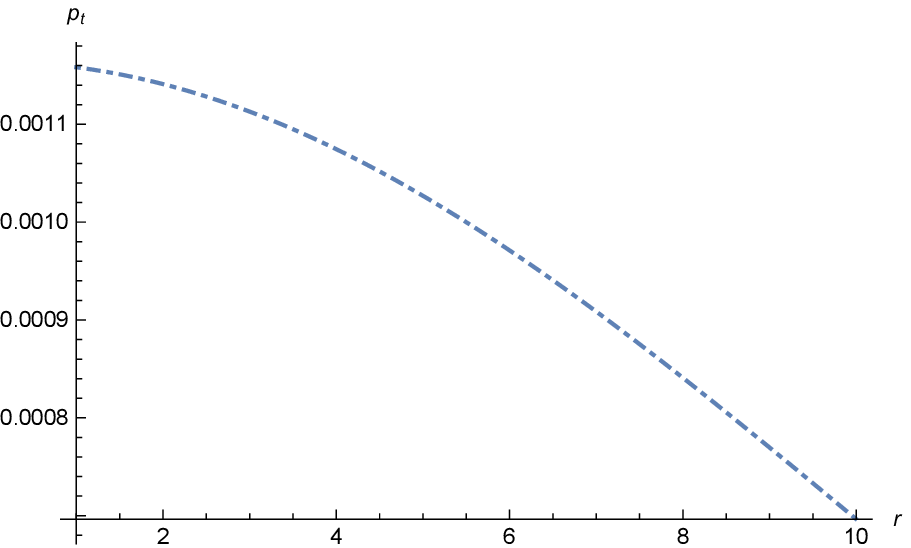,width=0.35\linewidth} &
\epsfig{file=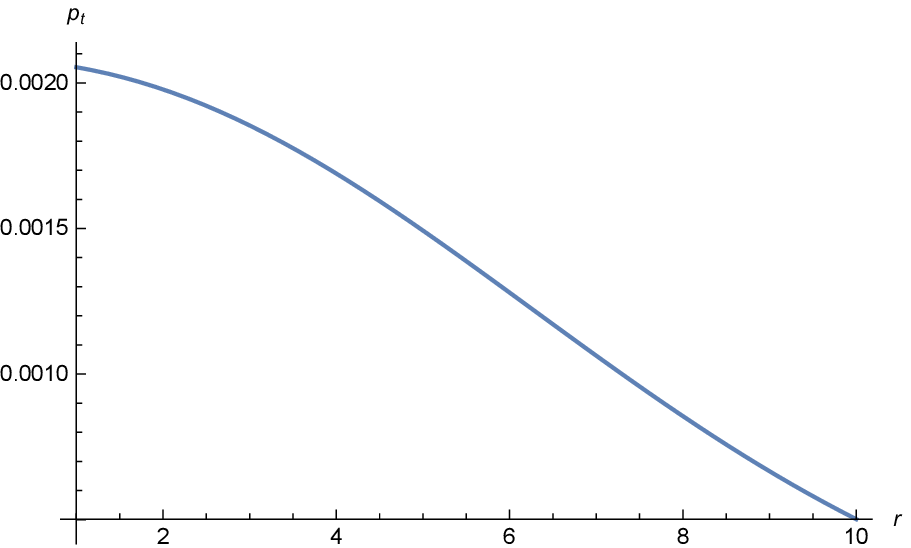,width=0.35\linewidth} &
\epsfig{file=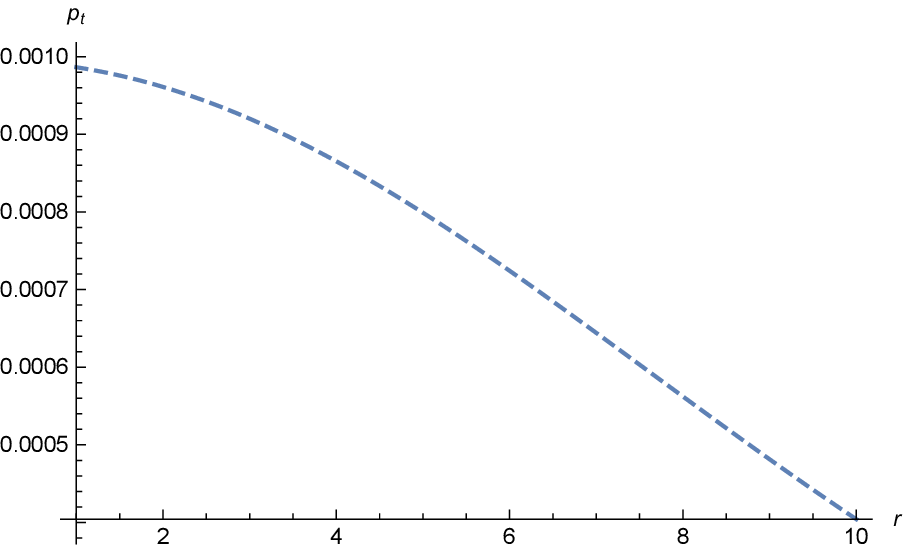,width=0.35\linewidth}\\
\end{tabular}
\caption{Plots of the transverse pressure evolution of the strange
star candidates Her X-1, SAX J 1808.4-3658, and 4U 1820-30;
represented by (left to right) 1st, 2nd, and 3rd graphs,
respectively.\label{fig:ptgraphs}}\center
\end{figure}
The plot of the energy density for the  strange star candidate Her X-1 (Figure \ref{fig:Densitygraphs})
shows that as $r\rightarrow0$, $\rho$ goes to maximum, and
this in fact indicates the high compactness of the core
of the star validating that our model under investigation is viable for the outer
region of the core. The other two graphs in the same row  provide us with similar sort of conclusions.
Furthermore, from the other plots of the anisotropic radial pressure $p_{r}$ (Figure $2$), it is evident that the radius
of the star for this model is $R=10$ km which further when used, gives us the density of $1.5828\times10^{15}$ g $cm^{-3}$,
a high value with a small radius of 10 km, showing that our $f(\mathcal{G},T)$ model is compatible to the structure of ultra-compact
star \cite{Ruderman,Herzog}. A comparison of this with the already existing data, labels this compact star as quark/strange star \cite{Bhar2}.
All the three plots in Figure $(3)$ indicate that the tangential pressure $ p_{t} $ remains positive and finite and show their decreasing behavior, which is required for the viability of the compact star model.
\begin{figure}\center
\begin{tabular}{cccc}
\\ &
\epsfig{file=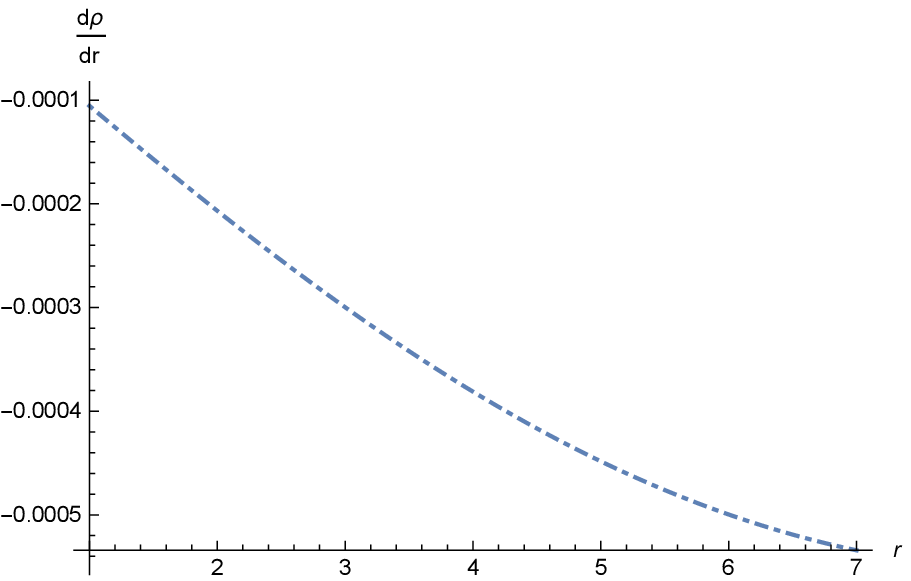,width=0.35\linewidth} &
\epsfig{file=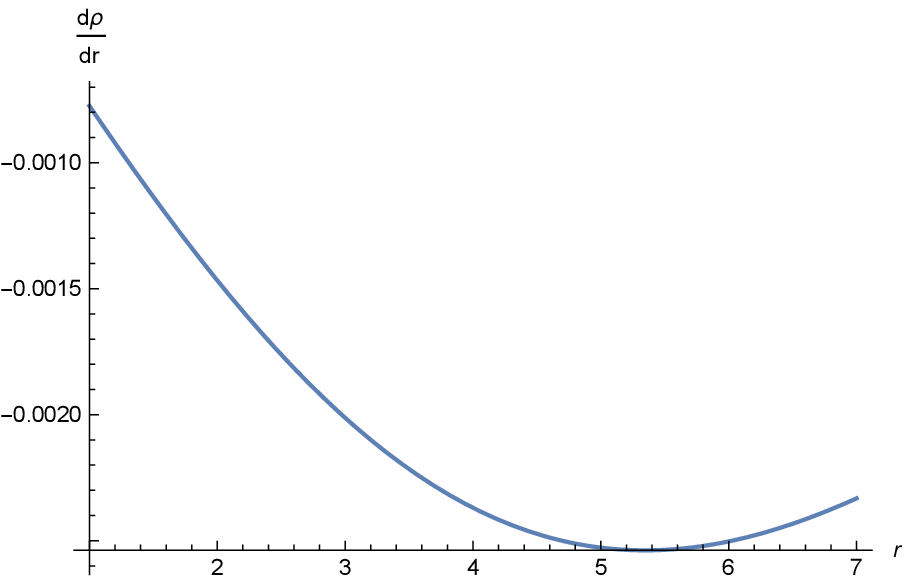,width=0.35\linewidth} &
\epsfig{file=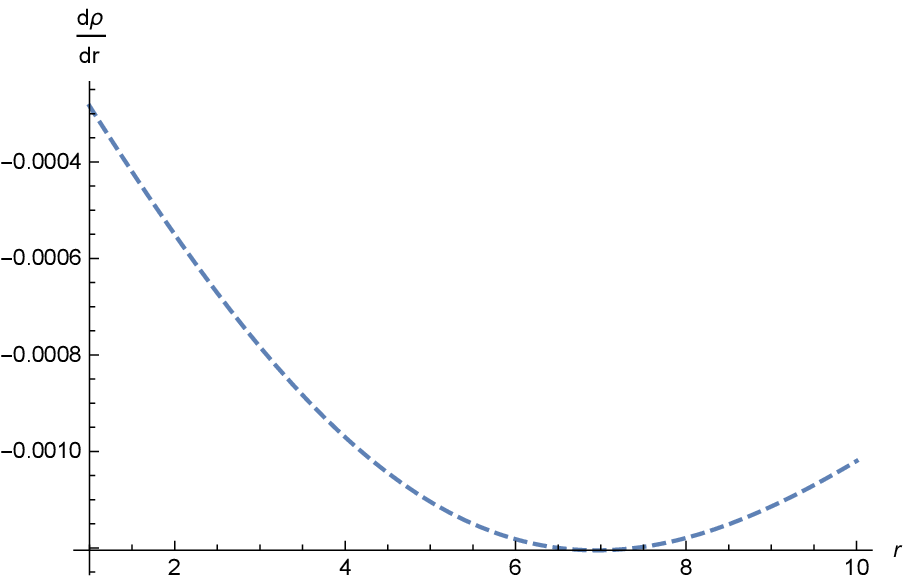,width=0.35\linewidth}\\
\end{tabular}
\caption{Plots of $\frac{d\rho}{dr}$ with increasing radius of the
strange star candidates Her X-1, SAX J 1808.4-3658, and 4U
1820-30; represented by (left to right) 1st, 2nd, and 3rd graphs,
respectively.\label{fig:Derivativerhographs}}\center
\end{figure}
\begin{figure}\center
\begin{tabular}{cccc}
\\ &
\epsfig{file=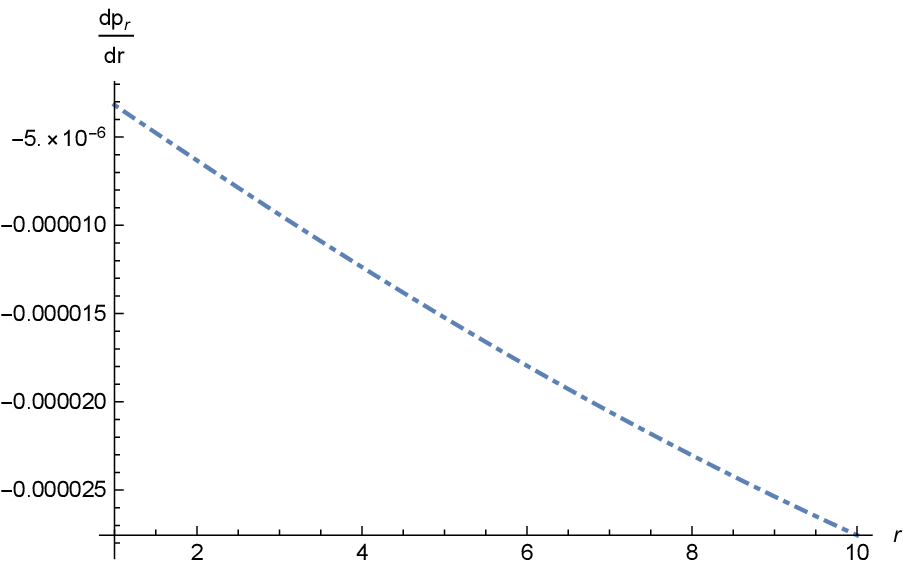,width=0.35\linewidth} &
\epsfig{file=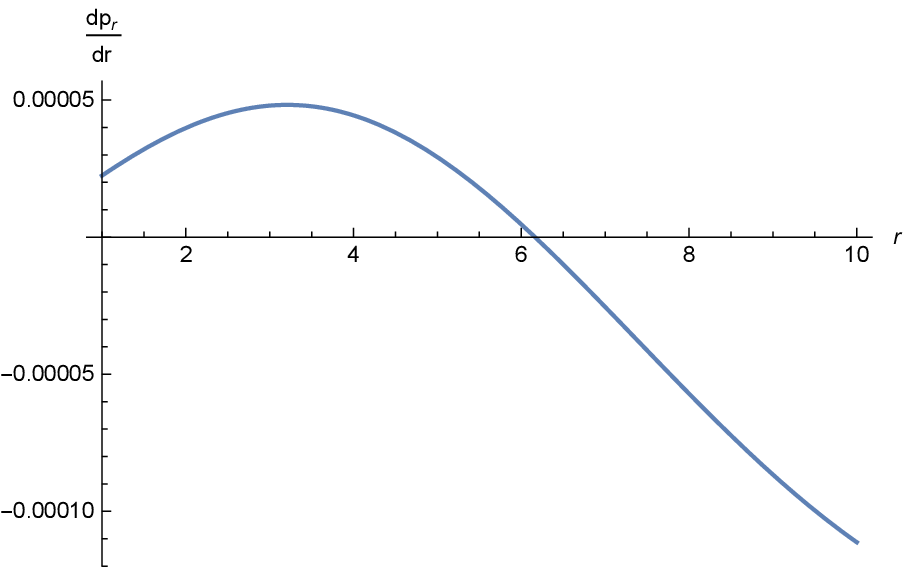,width=0.35\linewidth} &
\epsfig{file=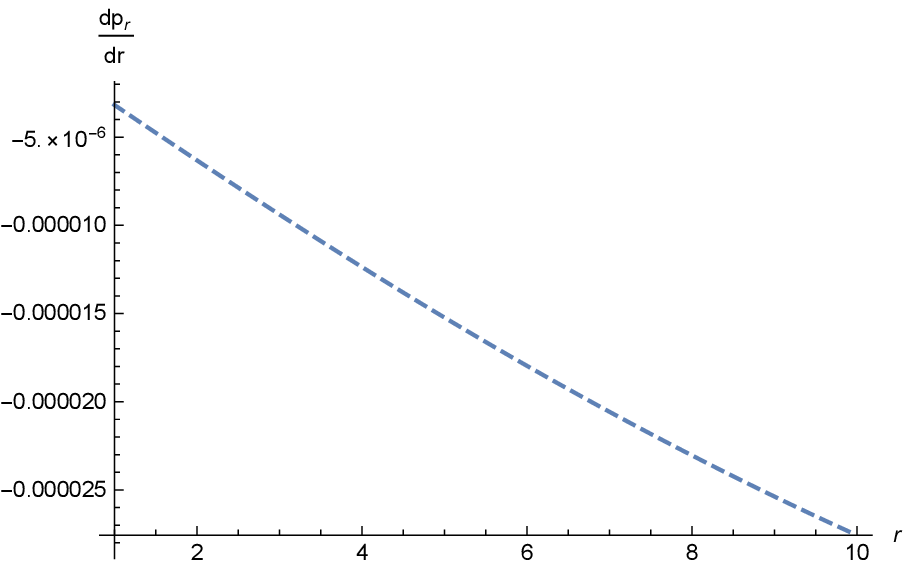,width=0.35\linewidth}\\
\end{tabular}
\caption{Variations of $\frac{dp_r}{dr}$ with increasing radius of
the strange star candidate Her X-1, SAX J 1808.4-3658, and 4U
1820-30; represented by (left to right) 1st, 2nd, and 3rd graphs,
respectively.\label{fig:Derivativeprgraphs}}\center
\end{figure}

The variations of the radial derivatives of the density and radial pressure are shown in Figures (\ref{fig:Derivativerhographs}) and (\ref{fig:Derivativeprgraphs}) respectively.
One can see the decreasing evolution of the first $r$-derivatives i.e $\frac{d\rho}{dr}<0$ and $\frac{dp_r}{dr}<0$. It may be noted here that at $r=0$,
these derivatives just disappear except of the radial pressure for the SAX J 1808.4-3658 candidate. 
The mathematical calculations for the second derivative test both for $\rho$ and $p_{r}$  tell us that $\frac{d^{2}\rho}{dr^{2}}<0$ and $\frac{d^{2}p_{r}}{dr^{2}}<0$, indicating the maximum values of the density and radial pressure at the center. This further suggests the compact nature of the star.
\subsection{Energy Conditions}
\begin{figure}\center
\begin{tabular}{cccc}
\\ &
\epsfig{file=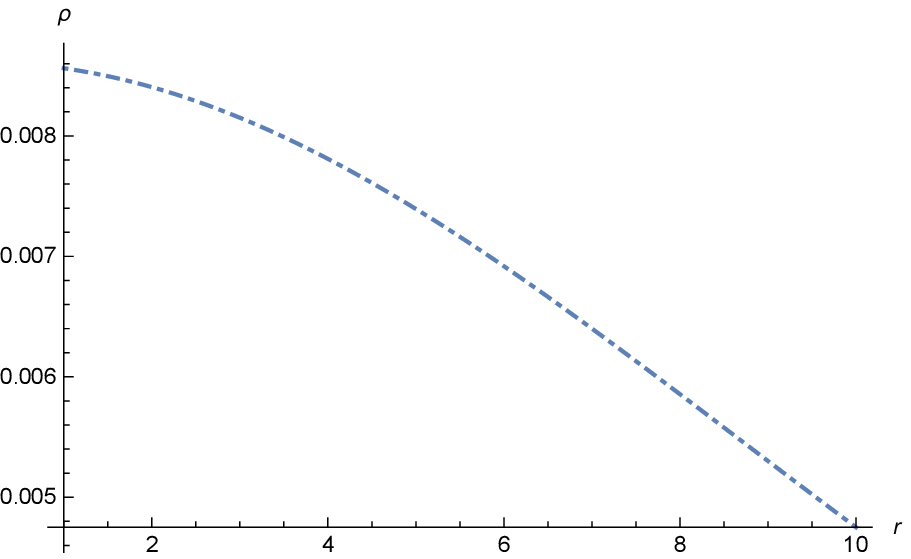,width=0.3\linewidth} &
\epsfig{file=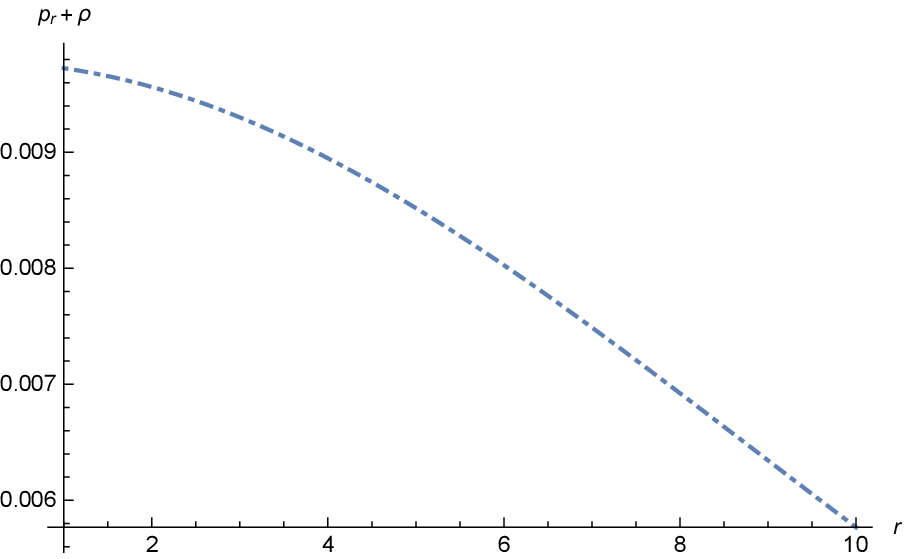,width=0.3\linewidth} &
\epsfig{file=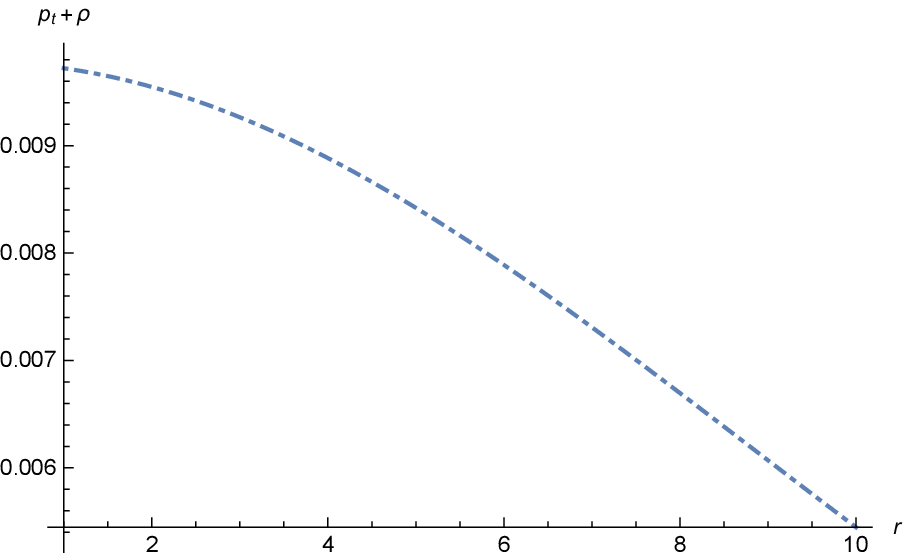,width=0.3\linewidth}\\&
\epsfig{file=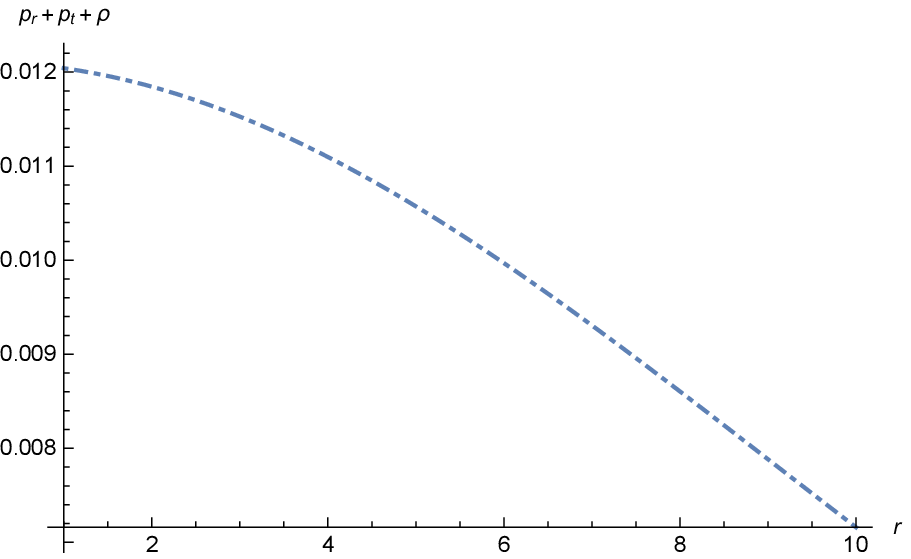,width=0.3\linewidth}&
\epsfig{file=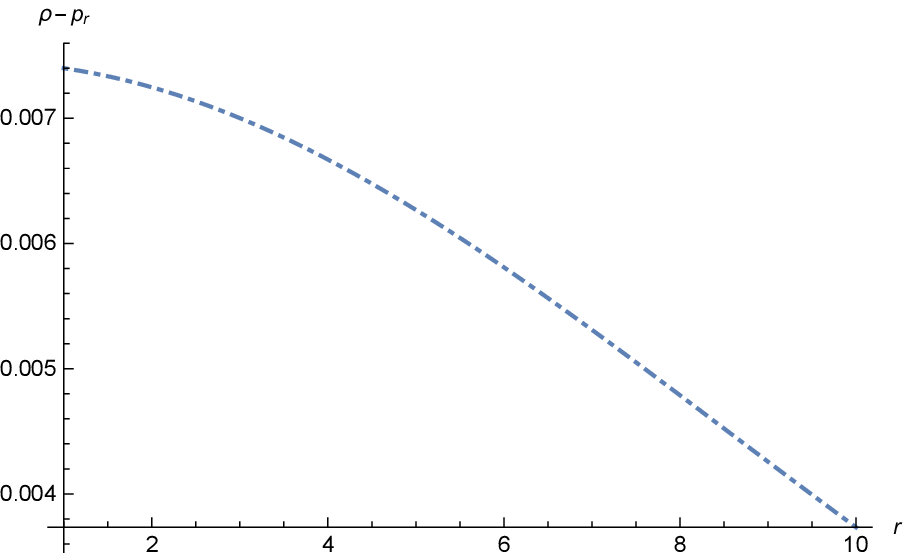,width=0.3\linewidth} &
\epsfig{file=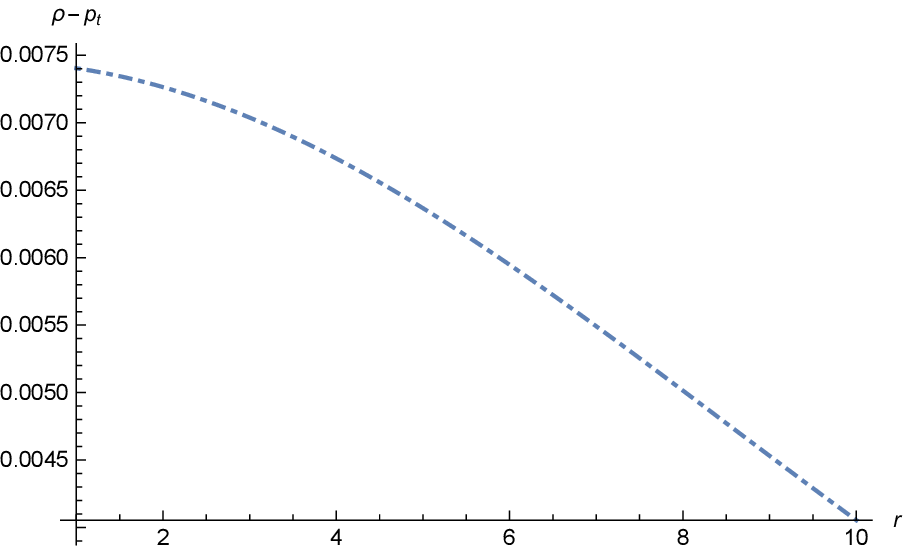,width=0.3\linewidth}\\
\end{tabular}
\caption{Plots of energy conditions  with respect to radius $r$ (km) of the strange star candidate Her X-1.\label{fig:ECs}}\center
\end{figure}
For the viability of the model, the energy bounds must be satisfied due to their significant importance in analysing the theoretical data. NEC (null energy conditions), WEC (weak energy conditions), SEC (strong energy conditions), and DEC (dominant energy conditions) have been given as
\begin{eqnarray}\nonumber
\textbf{NEC}&:& \rho+p_{r}\geq0,~~~~\rho+p_{t}\geq0,\\\nonumber
\textbf{WEC}&:& \rho\geq0,~~~~\rho+p_{r}\geq0,~~~~\rho+p_{t}\geq0,\\\nonumber
\textbf{SEC}&:& \rho+p_{r}\geq0,~~~~\rho+p_{t}\geq0,~~~~\rho+p_{r}+2p_{t}\geq0,\\\nonumber
\textbf{DEC}&:& \rho>|p_{r}|,~~~~\rho>|p_{t}|.
\end{eqnarray}
The evolution of all these energy conditions have been well satisfied as represented graphically for the  strange star candidate Her X-1 in Figure (\ref{fig:ECs}).
Hence our solutions are physically viable.
\subsection{Tolman-Oppenheimer-Volkoff (TOV) Equation}
\begin{figure}\center
\begin{tabular}{cccc}
\\ &
\epsfig{file=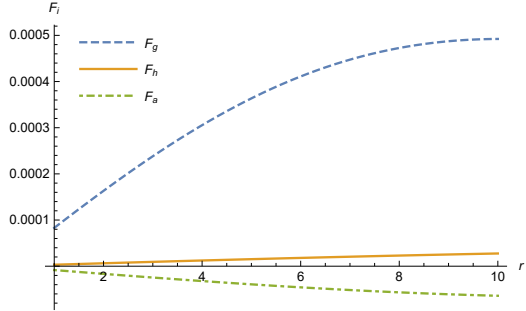,width=0.5\linewidth}\\
\end{tabular}
\caption{ The plot of gravitational force $(F_{g})$, hydrostatic
force $(F_{h})$ and anisotropic force $(F_{a})$
 for the  strange star candidate Her X-1  with respect to the radial coordinate $r$ (km).\label{fig:Eqlbrm}}\center
\end{figure}
In anisotropic case, the  generalized TOV equation is given as
\begin{eqnarray}\label{TOV}
\frac{M_{G}(r)(\rho+p_{r})}{r^{2}}e^{\frac{b-a}{2}}+\frac{dp_{r}}{dr}-\frac{2}{r}(p_{t}-p_{r})&=&0,
\end{eqnarray}
where
\begin{eqnarray}\label{MGr}
M_{G}(r)&=&\frac{1}{2}r^{2}e^{\frac{a-b}{2}}a'
\end{eqnarray} is the gravitational mass of a sphere of radius $r$. Now putting Eq.(\ref{MGr}) into Eq.(\ref{TOV}), it follows
\begin{eqnarray}\label{TOVE}
-\frac{a'}{2}(\rho+p_{r})-\frac{dp_{r}}{dr}+\frac{2}{r}(p_{t}-p_{r})&=&0.
\end{eqnarray}
Eq.(\ref{TOVE}) gives the information about the stellar configuration equilibrium under the combined effect of different forces like the anisotropic force $F_{a}$, the hydrostatic force $F_{h}$, and the gravitational force $F_{g}$. Their summation to zero eventually describes the equilibrium condition of the form
\begin{eqnarray}\label{Eqlbrumeqn}
F_{g}+F_{h}+F_{a}&=&0,
\end{eqnarray} where
\begin{eqnarray}
F_{g}&=&-\frac{a'}{2}(\rho+p_{r}),\\\nonumber
F_{h}&=&-\frac{dp_{r}}{dr},\\\nonumber
F_{a}&=&\frac{2}{r}(p_{t}-p_{r}).
\end{eqnarray}
From Figure (\ref{fig:Eqlbrm}), it can be noticed that under the
mutual effect of the three forces $F_{g}$, $F_{h}$ and $F_{a}$, the
static equilibrium might be achieved. It is mentioned here that at
some point if $p_{r}=p_{t}$ then $F_{a}=0$, which suggests that the
equilibrium becomes independent of the anisotropic force $F_{a}$.
\subsection{Stability Analysis}
\begin{figure}\center
\begin{tabular}{cccc}
\\ &
\epsfig{file=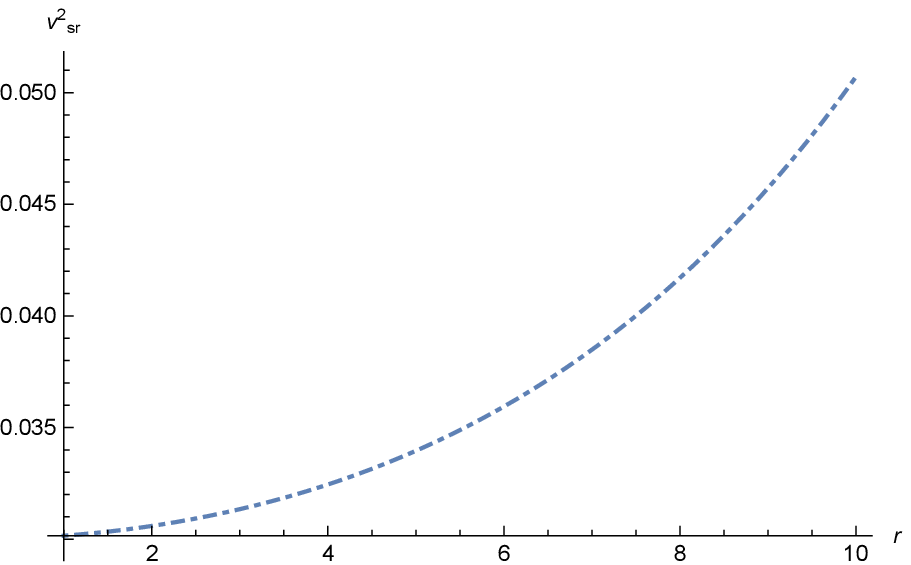,width=0.35\linewidth} &
\epsfig{file=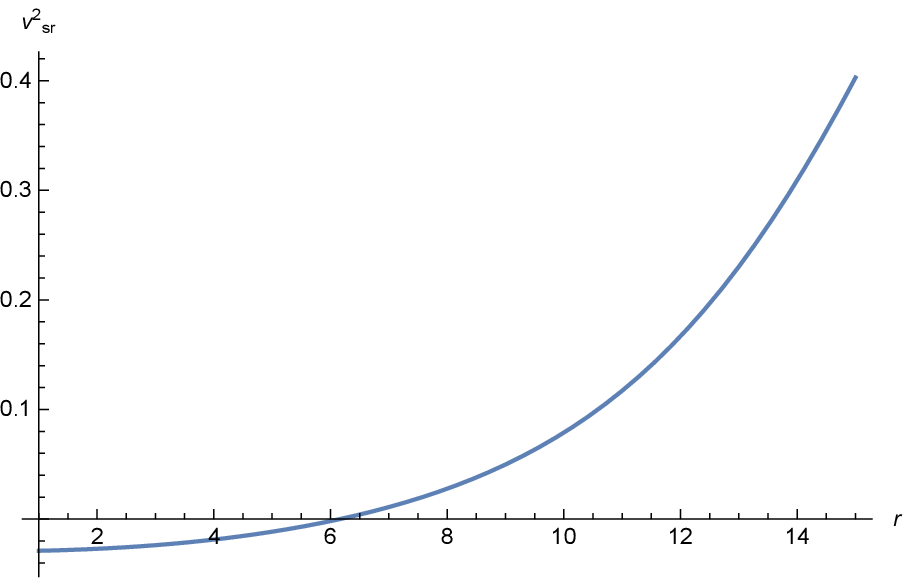,width=0.35\linewidth} &
\epsfig{file=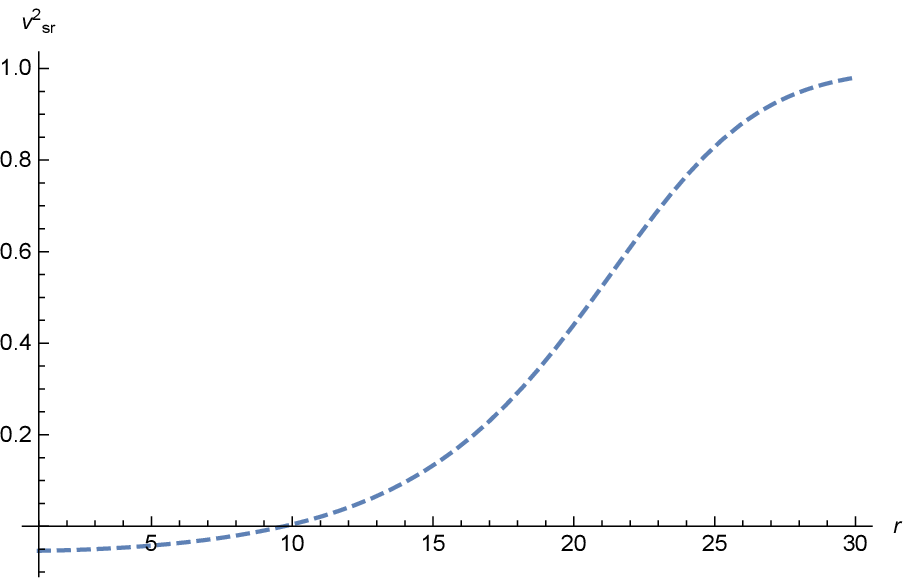,width=0.35\linewidth}\\
\end{tabular}
\caption{Variations of $v^{2}_{sr}$ with respect radius $r$ (km) of
the strange star candidate Her X-1, SAX J 1808.4-3658, and 4U
1820-30; represented by (left to right) 1st, 2nd, and 3rd graphs,
respectively.\label{fig:Vsr}}\center
\end{figure}
\begin{figure}\center
\begin{tabular}{cccc}
\\ &
\epsfig{file=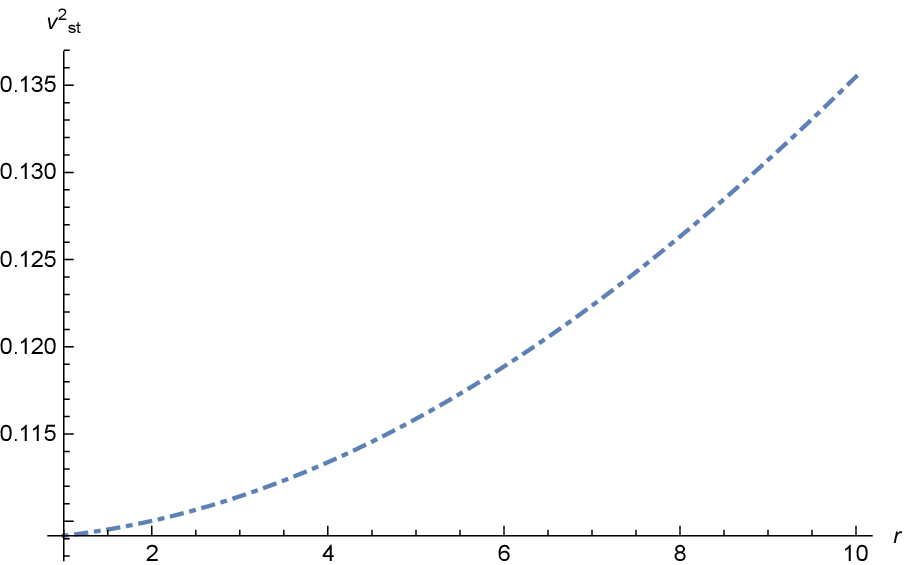,width=0.35\linewidth} &
\epsfig{file=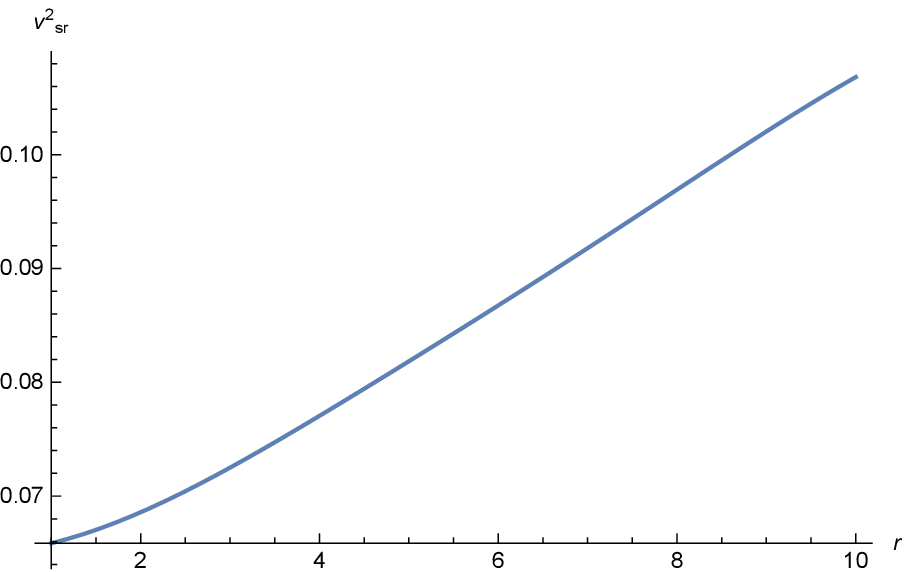,width=0.35\linewidth} &
\epsfig{file=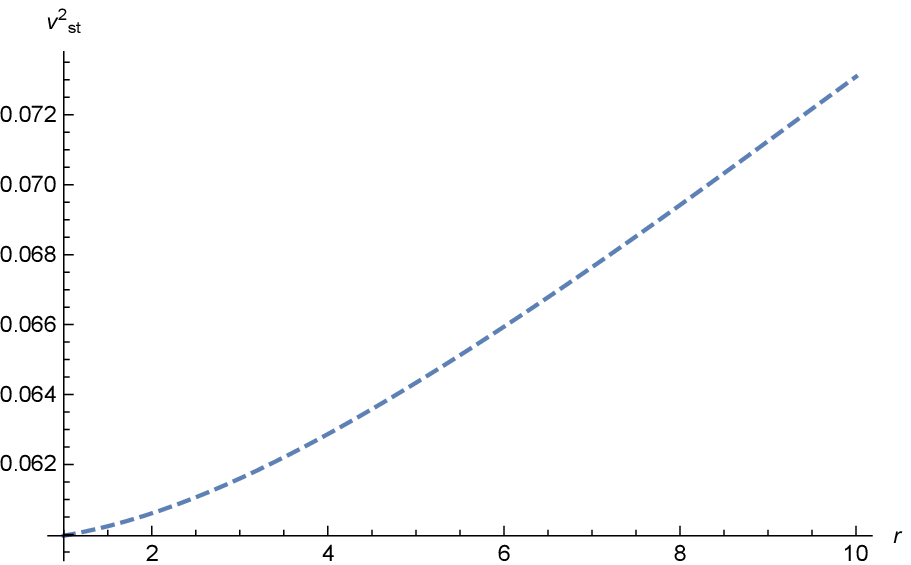,width=0.35\linewidth}\\
\end{tabular}
\caption{Variations of $v^{2}_{st}$ with respect radius $r$ (km) of
the strange star candidate Her X-1, SAX J 1808.4-3658, and 4U
1820-30; represented by (left to right) 1st, 2nd, and 3rd graphs,
respectively.\label{fig:Vst}}\center
\end{figure}

\begin{figure}\center
\begin{tabular}{cccc}
\\ &
\epsfig{file=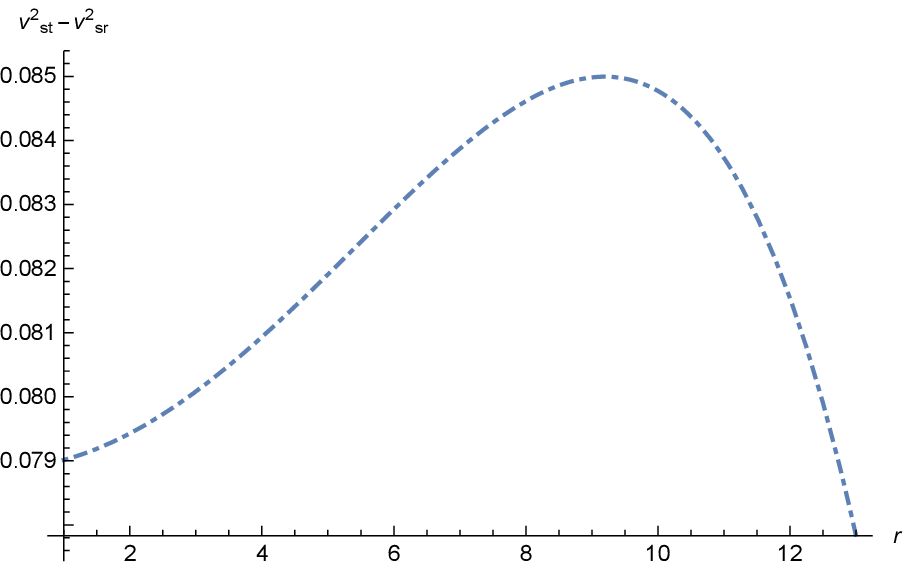,width=0.35\linewidth} &
\epsfig{file=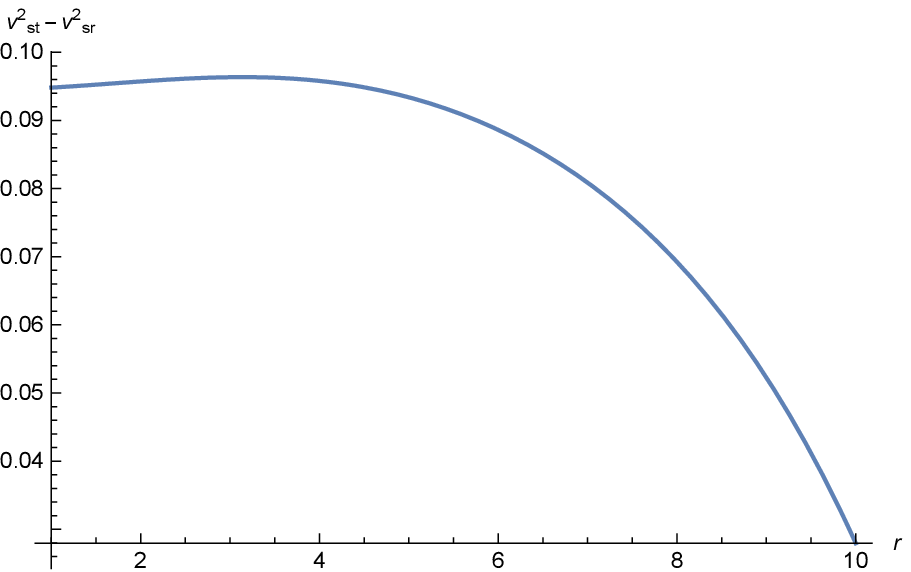,width=0.35\linewidth} &
\epsfig{file=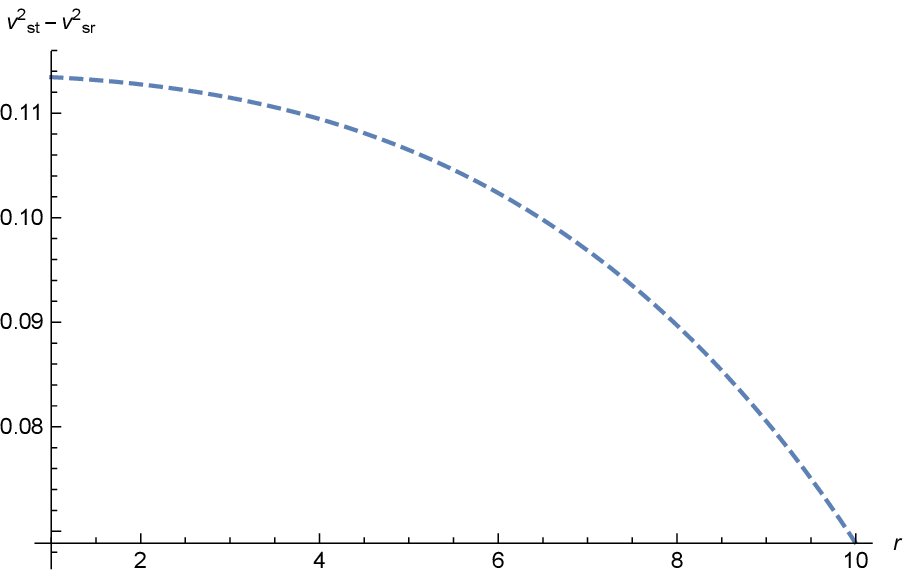,width=0.35\linewidth}\\
\end{tabular}
\caption{Variations of $v^{2}_{st}$ -$v^{2}_{sr}$ with respect to
radius $r$ (km) of the strange star candidate Her X-1, SAX J
1808.4-3658, and 4U 1820-30; represented by (left to right)
1st, 2nd, and 3rd graphs, respectively.\label{fig:Vst-Vsr}}\center
\end{figure}

For the stability, the radial and transversal sound speeds denoted
by $v^{2}_{sr}$ and $v^{2}_{st}$ respectively, should satisfy the
bounds, $0\leq{v^{2}_{sr}}\leq1$ and $0\leq{v^{2}_{st}}\leq1$
\cite{Herrera}, where $\frac{dp_{r}}{d\rho}=v^{2}_{sr}$ and
$\frac{dp_{t}}{d\rho}=v^{2}_{st}$. It can be seen from the Figures
(\ref{fig:Vsr}) and (\ref{fig:Vst}) that the evolution of the radial
and transversal sound speeds for strange star candidate Her X-1 are
within the bounds of stability as discussed, but in the case
of SAX J 1808.4-3658, and 4U 1820-30 strange star candidates,
the radial sound speeds evolution which is against the radial
coordinate $r$ temporarily violates these stability conditions.
However, for the same candidates the transversal sound speeds are
satisfied. Within the matter distribution, the estimation of the
potentially stable and unstable eras can be had from the differences
of the propagations of the sound speeds which is the expression
$v^{2}_{st}-v^{2}_{sr}$ satisfying the inequality
$0<|v^{2}_{st}-v^{2}_{sr}|<1$. This can be seen clearly from the
plots of Figure (\ref{fig:Vst-Vsr}). Thus, overall the stability may be 
attained for compact stars under $f(\mathcal{G},T)$ gravity model, particularly for
the  strange star candidate Her X-1.

\subsection{Adiabatic Index Analysis}

For the case of anisotropic fluid spherical star, as proposed in \cite{Chandrasekhar,Heintzmann}, the stability depends on the adiabatic index $\gamma$, and for the radial and tangential cases, we respectively have
\begin{equation}
\gamma_{r}=\frac{\rho+p_{r}}{p_{r}}\frac{dp_{r}}{d\rho}~~~~\text{and}~~~~\gamma_{t}=\frac{\rho+p_{t}}{p_{t}}\frac{dp_{t}}{d\rho}.
\end{equation}
The stability of a Newtonian sphere should be satisfied if $\gamma>\frac{4}{3}$, and $\gamma=\frac{4}{3}$ is the condition for the occurrence of neutral equilibrium \cite{Bondi}. Due to the presence of the effective pressure, the anisotropic relativistic sphere obeys more complicated stability condition given as
\begin{equation}
\gamma>
\frac{4}{3}+\Big[\frac{4}{3}\frac{p_{t_{0}}-p_{r0}}{|p'_{r_{0}}|r}+\frac{8\pi}{3}\frac{\rho_{0}p_{r_{0}}r}{|p'_{r_{0}}|}\Big]_{max},
\end{equation}
where $\rho_{0}$, $p_{t_{0}}$, and $p_{r_{0}}$ are the initial energy density, tangential pressure, and radial pressure in static equilibrium respectively satisfying Eq. (\ref{TOV}). In our case, the adiabatic index has been calculated analytically for strange star candidate Her X-1 , giving us $\gamma_{r}=1.3757$ and $\gamma_{t}=1.0597$, which shows the complete stability in radial case but some deviation in the tangential case.

\subsection{Mass-Radius Relationship}

\begin{figure}\center
\begin{tabular}{cccc}
\\ &
\epsfig{file=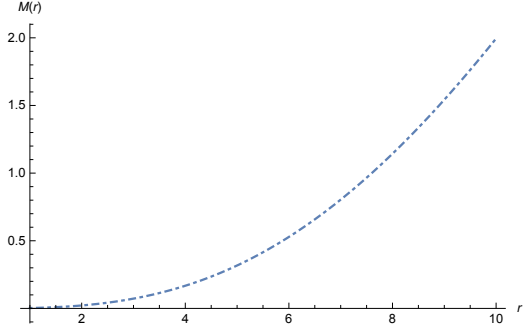,width=0.5\linewidth}\\
\end{tabular}
\caption{ Evolution of mass-function $M(r)$ for the strange star candidate Her X-1 with respect to the radial coordinate $ r$ (km).\label{fig:massfunction}}\center
\end{figure}
The mass of the compact star as a function of radius $r$ is given as
\begin{equation}\label{mass}
M(r)=4\int_{0}^{r}\pi\acute{r^2}\rho{d}\acute{r}.
\end{equation}
It can be seen clearly from the profile of the mass function given
in Figure (\ref{fig:massfunction}) that the mass of the star is
directly proportional to the radius, and $M(r)\rightarrow0$ as
$r\rightarrow0$, which depicts that the mass function is regular at
the centre of the star. Moreover, for the spherically symmetric
anisotropic perfect fluid case, the ratio of the mass to the radius,
according to Buchdahl \cite{Buchdahl} should be bounded like
$\frac{2M}{r}\leq\frac{8}{9}$. In our case, the situation is very good as we get $\frac{2M}{r}=0.4987$
and the condition is clearly satisfied.

\subsection{ Compactness and Redshift Analysis}

\begin{figure}\center
\begin{tabular}{cccc}
\\ &
\epsfig{file=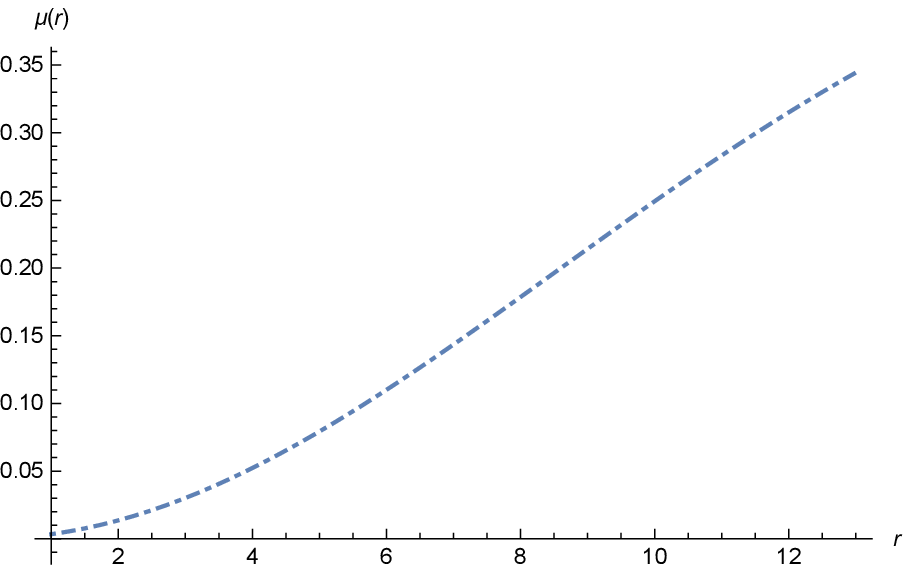,width=0.4\linewidth} &
\epsfig{file=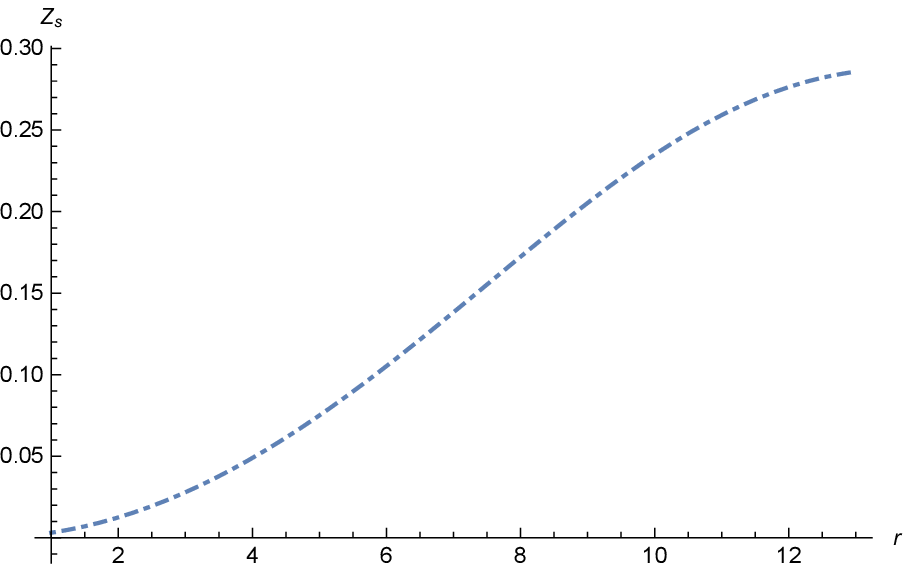,width=0.4\linewidth}\\
\end{tabular}
\caption{ Plots of the Compactness $\mu(r)$ and Surface redshift $(Z_{s})$  of the strange star candidate Her X-1 with respect to the radial coordinate $r$ (km).\label{fig:redshift}}\center
\end{figure}
Compactness $\mu(r)$ of the star is expressed as
\begin{equation}\label{Compactness}
\mu(r)=\frac{4}{r}\int_{0}^{r}\pi\acute{r^2}\rho{d}\acute{r}.
\end{equation}
Therefore, the redshift $Z_{S}$ is determined as
\begin{equation}\label{RedShift}
Z_{S}+1=[-2\mu(r)+1]^{\frac{-1}{2}}.
\end{equation}
The graphical evolution of the surface redshift has been given in
Figure (\ref{fig:redshift}). The value of the function $Z_{S}$ for the case of the strange star candidate Her X-1 is calculated as  $Z_{S}\approx0.22$ which is within the desired bound of $Z_{S}\leq2$. 
\begin{figure}\center
\begin{tabular}{cccc}
\\ &
\epsfig{file=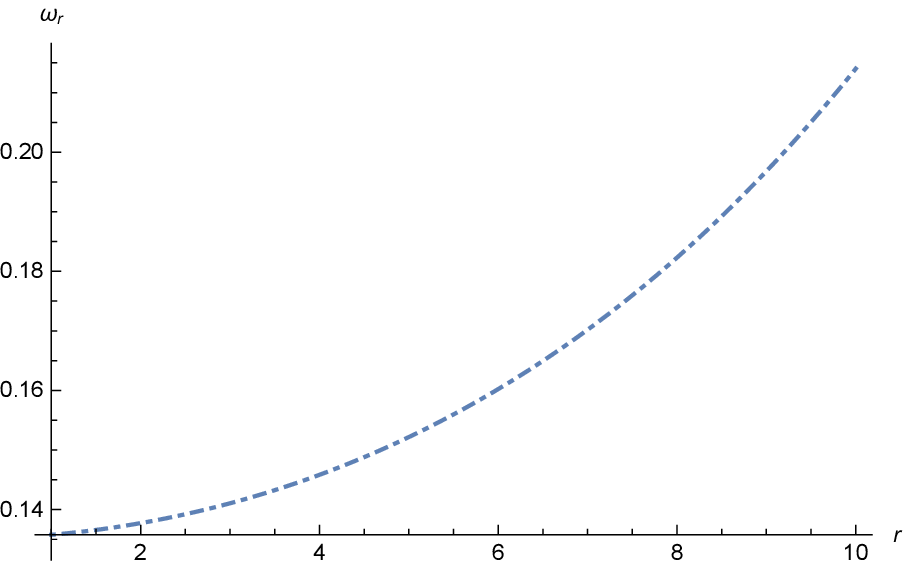,width=0.35\linewidth} &
\epsfig{file=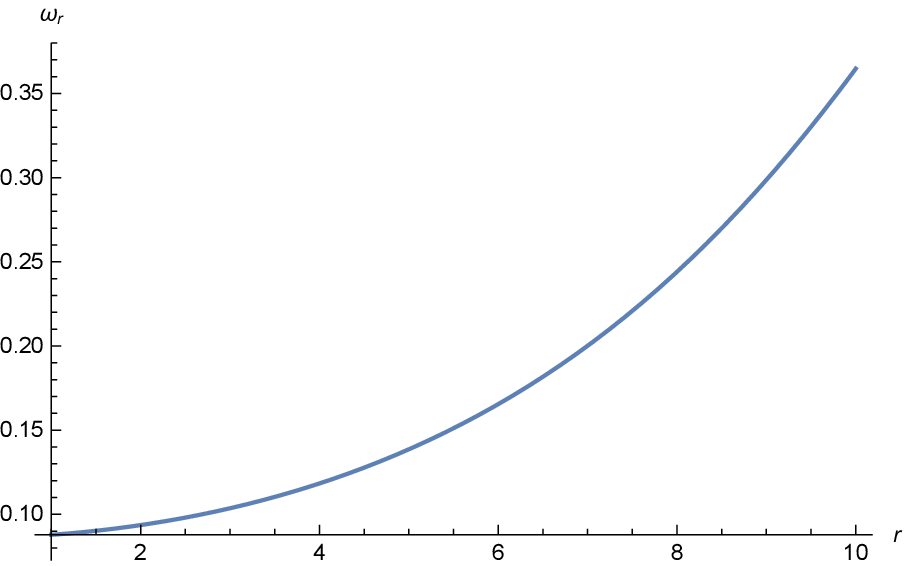,width=0.35\linewidth} &
\epsfig{file=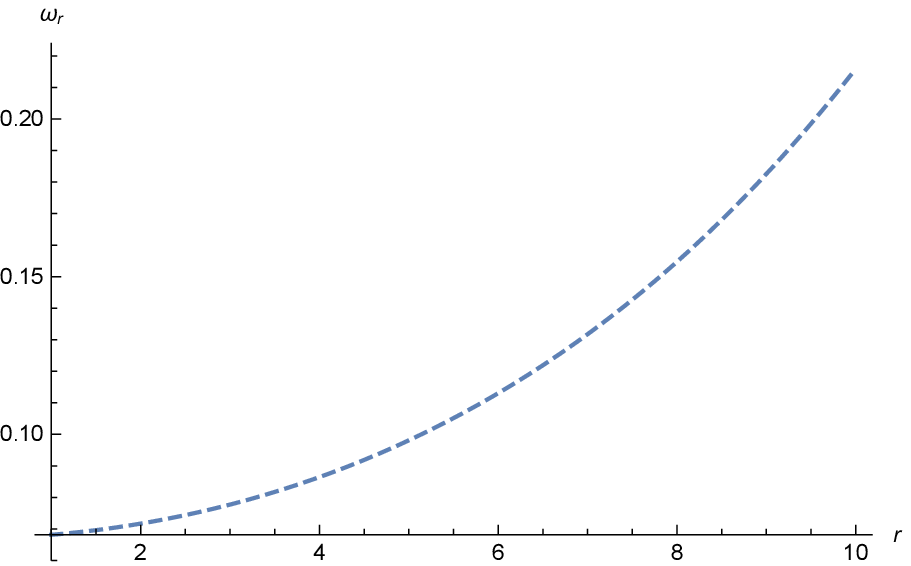,width=0.35\linewidth}\\
\end{tabular}
\caption{Variations of the radial EoS parameter with respect to the
radial coordinate $r$(km) of the strange star candidate Her X-1, SAX
J 1808.4-3658, and 4U 1820-30; represented by (left to right)
1st, 2nd, and 3rd graphs, respectively.\label{fig:rEoS}}\center
\end{figure}
\begin{figure}\center
\begin{tabular}{cccc}
\\ &
\epsfig{file=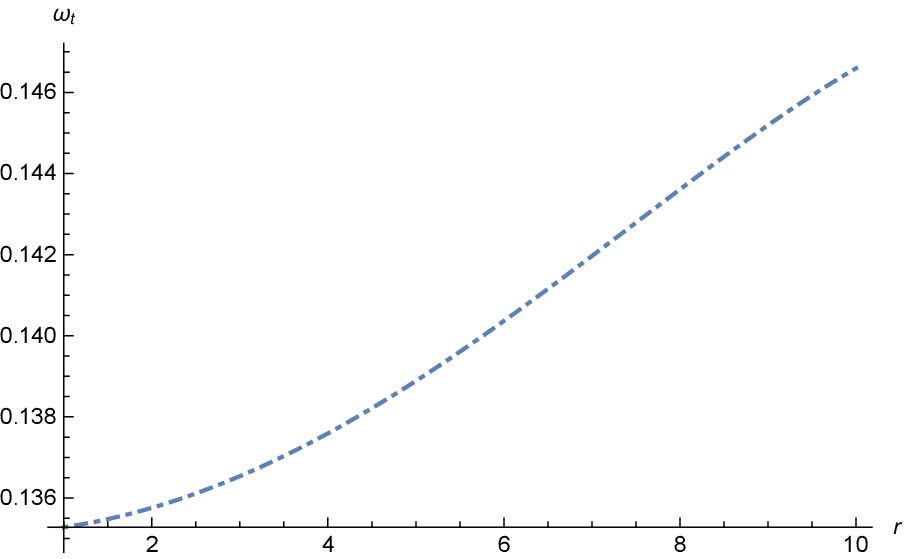,width=0.35\linewidth} &
\epsfig{file=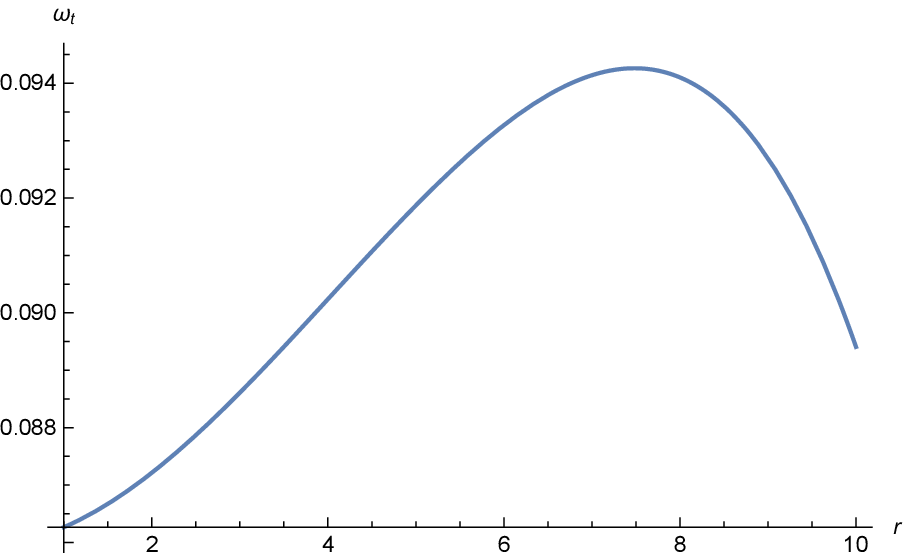,width=0.35\linewidth} &
\epsfig{file=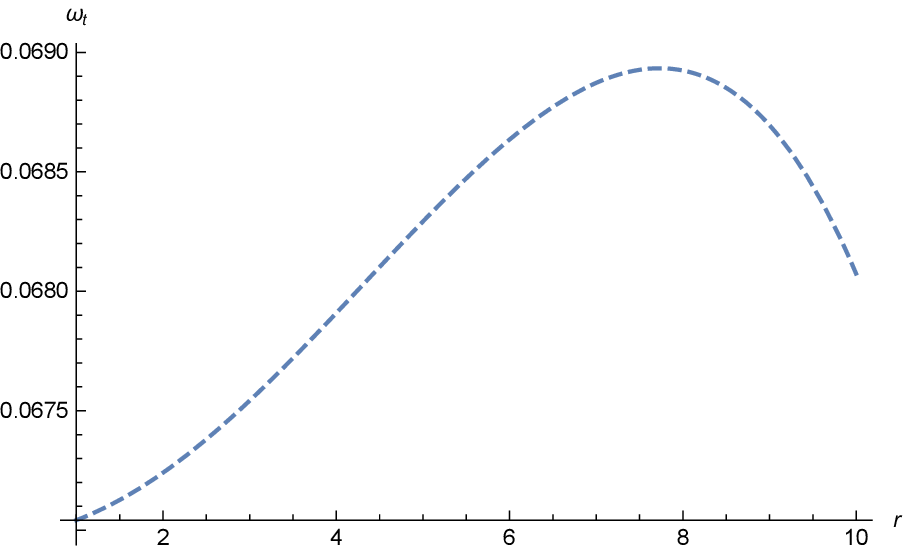,width=0.35\linewidth}\\
\end{tabular}
\caption{Variations of the tangential EoS parameter with respect to
the radial coordinate $r$ (km) of the strange star candidate Her
X-1, SAX J 1808.4-3658, and 4U 1820-30; represented by (left
to right) 1st, 2nd, and 3rd graphs,
respectively.\label{fig:tEoS}}\center
\end{figure}

\subsection{EoS Parameter and the Measurement of Anisotropy}

Now for anisotropic case, the
radial and transversal forms of EoS parameter can be written as
\begin{equation}
\omega_{r}=\frac{p_{r}}{\rho}~~~~\text{and}~~~~\omega_{t}=\frac{p_{t}}{\rho}.
\end{equation}
The evolution of these EoS parameters with the increasing radius
have been shown in Figures (\ref{fig:rEoS} and \ref{fig:tEoS}) which
clearly demonstrate that all the six plots satisfy the inequalities
$0<\omega_{r}<1$ and $0<\omega_{t}<1$. This further advocates the effectiveness of the
considered model.
\begin{figure}\center
\begin{tabular}{cccc}
\\ &
\epsfig{file=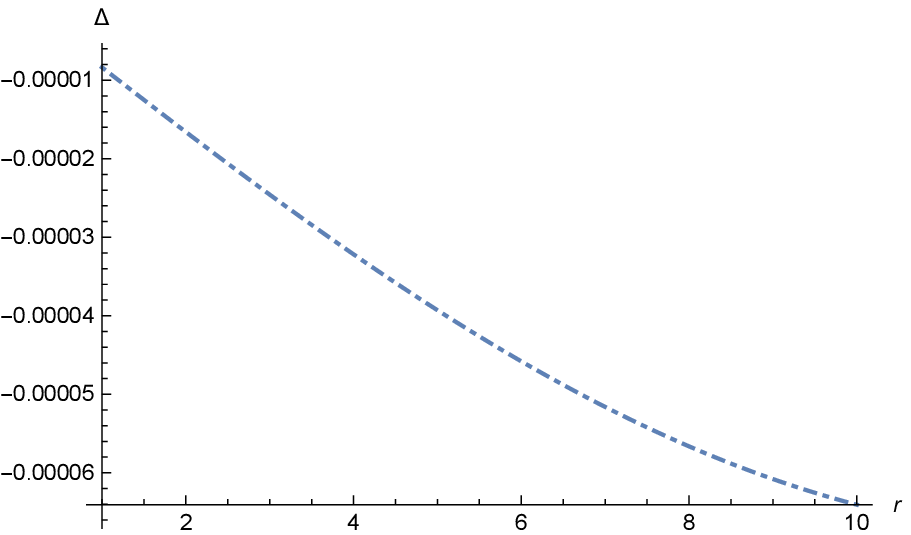,width=0.35\linewidth} &
\epsfig{file=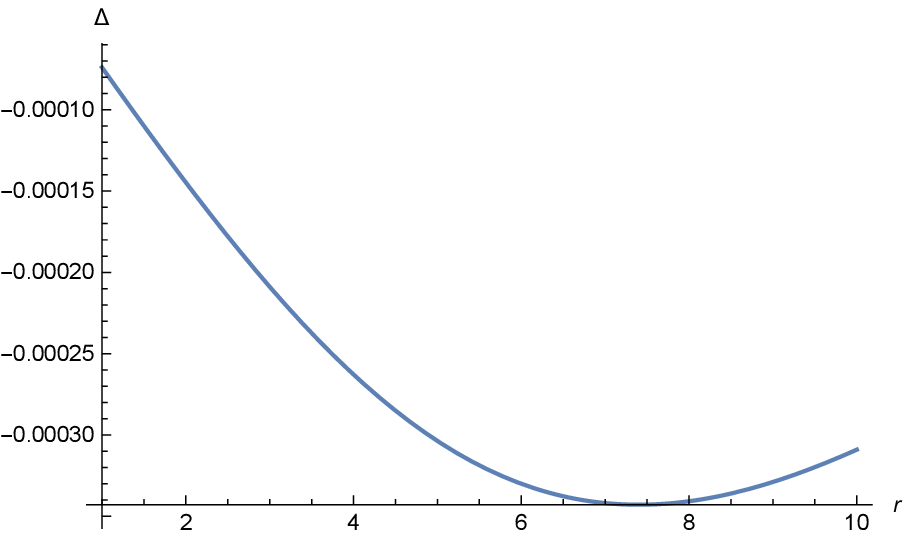,width=0.35\linewidth} &
\epsfig{file=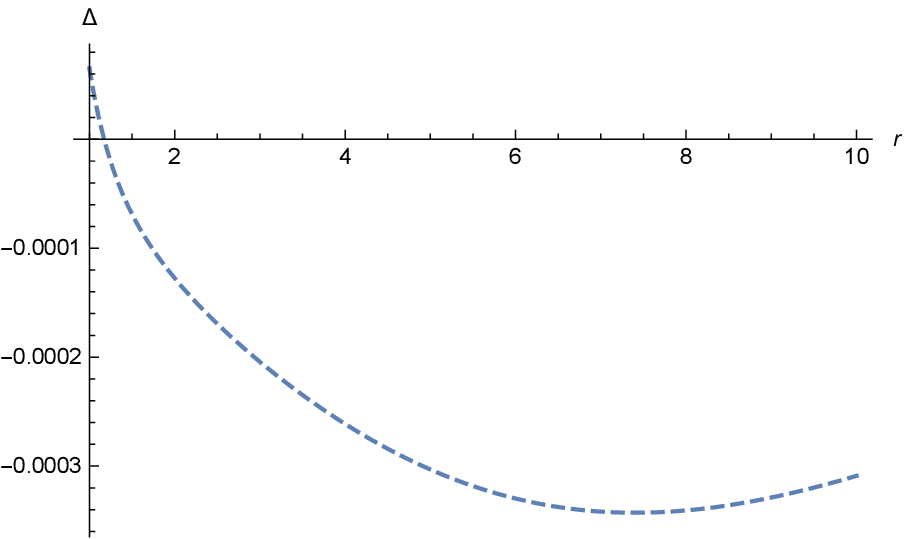,width=0.35\linewidth}\\
\end{tabular}
\caption{Variations of anisotropic measure $\Delta$ with respect to
the radial coordinate $r$ (km) of the strange star candidate Her
X-1, SAX J 1808.4-3658, and 4U 1820-30; represented by (left
to right) 1st, 2nd, and 3rd graphs,
respectively.\label{fig:Delta}}\center
\end{figure}

The measurement of the anisotropy denoted by $\Delta$ is given by
\begin{equation}\label{anisotropy}
\Delta=\frac{2}{r}{(p_t-p_r)},
\end{equation}
which gives the information about the anisotropic behavior of the
model. The $\Delta$ remains positive if $p_t>p_r$, suggesting the
anisotropy being drawn outward, and for the reverted situation,
i.e., $p_t<p_r$, the anisotropy  $\Delta$ turns negative which
corresponds to being directed inward. For our situation, the
variations of the anisotropic measurement $\Delta$ with respect to
the radial coordinate $r$ show the decreasing negative behavior
for the strange star candidate Her X-1 and SAX J 1808.4-3658
suggesting that $p_t<p_r$. For the 4U 1820-30 candidate, it remains
positive for a fraction of the radius $r$ where some repulsive
anisotropic force followed by massive matter distribution  appears and very soon, it gets negative after $r=0.25$.

\section{Concluding Discussion}

In this paper, we have put some useful discussions related to the emergence of
compact stars in the newly introduced $f(\mathcal{G},T)$ theory of gravity by
considering the model
$f(\mathcal{G},T)=\alpha\mathcal{G}^n+\lambda{T}$. We have tested
this model for the strange star candidates Her X-1, SAX J
1808.4-3658, and 4U 1820-30 for anisotropic case by using the Krori and Barua approach of metric function \cite{K&B}, that is,
$a=Br^2+C,~b=Ar^2$. The arbitrary constants $A$, $B$, and $C$ are
calculated by smoothly matching the interior metric conditions with
the Schwarzschild's exterior metric conditions. This phenomenon
makes us to understand the nature of the compact stars by expressing
their masses and  radii in terms of the arbitrary constants.

By using these constants in our investigation for the strange star
candidates Her X-1, SAX J 1808.4-3658, and 4U 1820-30 the
energy density, radial and tangential pressures have been plotted with
respect to radial coordinate $r$ indicating that when $r$
approaches to zero, the density goes to its maximum for all the
three strange star candidates. The same is the situation for the
tangential pressure but different behavior in the case of the radial
pressure for the 4U 1820-30 candidate. Mainly, this situation
admits the theory that the core of compact stars under
consideration, is intensely compact, particularly in case of the
strange star candidate Her X-1. We have succeeded to determine
the density of the emerging compact star $(\approx
1.5828\times10^{15}~g~cm^{-3})$ after estimating the radius of the
star $(R\approx10)$ from the evolution of the radial pressure. The
evolution of EoS parameters with the increasing radius satisfies the
inequalities $0<\omega_{r}<1$ and $0<\omega_{t}<1$ for the radial
and tangential EoS parameters respectively, which favor the
acceptance of the model under study.
We have also shown through the graphical representation that all the energy conditions namely NEC, WEC, SEC, and DEC are satisfied for the proposed
$f(\mathcal{G},T)$ gravity model in the case of Her X-1 favoring the physical viability of the model.

The static equilibrium, to some extent, has been established by plotting the three forces $F_{g}$, $F_{h}$, and $F_{a}$ comprised in the TOV equation.
The evolution of the radial and transversal sound speeds denoted by $v^{2}_{sr}$ and $v^{2}_{st}$ respectively for strange star candidate Her X-1 are within the bounds of stability \cite{Herrera} , but in the case of SAX J 1808.4-3658, and 4U 1820-30 strange star candidates, the radial sound speed $v^{2}_{sr}$ evolutions temporarily violate these stability conditions. However, for the same candidates the transversal sound speeds $v^{2}_{st}$ are satisfied. Within the matter distribution, the estimation of the strongly stable and unstable eras from the differences of the propagations of the sound speeds satisfies the inequality $0<|v^{2}_{st}-v^{2}_{sr}|<1$ for all the candidates as shown in Figure $11$. Thus, overall the stability is attained for compact star $f(\mathcal{G},T)$ model, particularly for the  strange star candidate Her X-1.
   
We have also investigated the dynamical stability by analytically calculating the adiabatic index $\gamma$ of the model both for the radial and tangential pressures  for strange star candidate Her X-1, giving us $\gamma_{r}=1.3757>4/3$ and $\gamma_{t}=1.0597$, which shows the complete stability in radial case but a slight deviation in the tangential case. We have found the direct proportionality of the mass function to the radius, and $M(r)\rightarrow0$ as $r\rightarrow0$, suggesting that the mass function is regular at the center of the star. Moreover, for the spherically symmetric anisotropic fluid case, the ratio of the mass to the radius has been calculated as $\frac{2M}{r}=0.4987$ satisfying $\frac{2M}{r}\leq\frac{8}{9}$ as proposed by Buchdahl \cite{Buchdahl}. The evolution of the compactness of the star for Her X-1 favors the model. The values of surface redshift function $Z_{S}$ are within the bound of $Z_{S}\leq2$ and for the case of the strange star candidate Her X-1, the redshift value is calculated as $Z_{S}\approx0.22$ which satisfies the upper bound $Z_{S}\leq2$. This further indicates the stability of the model under study. Conclusively, for the case of the strange star candidate Her X-1, all the physical parameters have been more consistent to favor the $f(\mathcal{G},T)$ gravity model under study as compared to the other two candidates SAX J 1808.4-3658 and 4U 1820-30. The overall consistency for the model may be improved by considering some more suitable choices of the physical parameters.\\\\
\textbf{Acknowledgement}\\\\ Many thanks to the anonymous reviewer
for valuable comments and suggestions to improve the paper.
This work was supported by National University
of Computer and Emerging Sciences (NUCES).
%%%%%%%%%%%%%%%%%%%%%%%%%%%%%%%%%%%%%%%%%%%%%%%%%%%%%%%%%%%%%%%%%%%%%%%%%%%%%%%%%%%%%%%%%%%%%%%%%%%%%%%%%%

\end{document}